\documentclass[12pt,preprint2]{aastex6}

\newcommand{\Ha}{H$\alpha$}

\newcommand{\Hb}{H$\beta$}
\newcommand{\cHb}{c$_{H\beta}$}
\newcommand{\Htwo}{H\textsubscript{2}}

\newcommand{\heI}{\ion{He}{1}}

\newcommand{\kms}{km~s{$^{-1}$}}

\newcommand{\cmq}{cm$^{-3}$}

\newcommand{\Hplus}{H$^{+}$}
\newcommand{\Heplus}{He$^{+}$}
\newcommand{\ozone}{He$^{+}$+H$^{+}$}

\newcommand{\nzone}{He$\rm ^{o}$+H$^{+}$}

\newcommand{\tC}{{$\theta^1$~Ori~C}}
\newcommand{\tB}{{$\theta^1$~Ori~B}}
\newcommand{\tA}{{$\theta^2$~Ori~A}}
\newcommand{\tAt}{{$\theta^1$~Ori~A}}
\newcommand{\tD}{{$\theta^1$~Ori~D}}

\newcommand{\hii}{\ion{H}{2}}
\newcommand{\hi}{\ion{H}{1}}

\newcommand{\oiii}{[\ion{O}{3}]}
\newcommand{\oii}{[\ion{O}{2}]}
\newcommand{\Cii}{[\ion{C}{2}]}
\newcommand{\oi}{[\ion{O}{1}]}

\newcommand{\nii}{[\ion{N}{2}]}

\newcommand{\siii}{[\ion{S}{3}]}
\newcommand{\Caii}{\ion{Ca}{2}}
\newcommand{\Nai}{\ion{Na}{1}}
\newcommand{\siiia}{\ion{S}{3}}
\newcommand{\Piii}{\ion{P}{3}}

\newcommand{\Vscat}{{\bf V$\rm_{scat}$}}
\newcommand{\Vscatnii}{{\bf V$\rm_{scat,[N II]}$}}
\newcommand{\Vscatoiii}{{\bf V$\rm_{scat,[O III]}$}}
\newcommand{\Vredoiii}{{\bf V$\rm_{red,[O III]}$}}

\newcommand{\Vblue}{{\bf V$\rm_{blue}$}}
\newcommand{\Vbluenii}{{\bf V$\rm_{blue,[N II]}$}}
\newcommand{\Vblueoiii}{{\bf V$\rm_{blue,[O III]}$}}
\newcommand{\Vlow}{{\bf V$\rm _{low}$}}

\newcommand{\Vlownii}{{\bf V$\rm _{low,[N II]}$}}
\newcommand{\Vlowoiii}{{\bf V$\rm _{low,[O III]}$}}

 \newcommand{\Vnewoiii}{{\bf V$\rm _{new,[O~III]}$}}
\newcommand{\Vnewnii}{{\bf V$\rm _{new,[N~II]}$}}
\newcommand{\Vnew}{{\bf V$\rm _{new}$}}
\newcommand{\Vpdr}{{\bf V$\rm _{PDR}$}}

\newcommand{\Vmifnii}{{\bf V$\rm_{mif,[N~II]}$}}
\newcommand{\Vmifoiii}{{\bf V$\rm_{mif,[O~III]}$}}

\newcommand{\Vmif}{{\bf V$\rm_{mif}$}}

\newcommand{\Voiii}{{\bf V$\rm_{[O~III]}$}}
\newcommand{\Vcii}{{\bf V$\rm_{[C~II]}$}}
\newcommand{\Vevap}{{\bf V$\rm_{evap}$}}
\newcommand{\Vevapnii}{{\bf V$\rm_{evap,[N~II]}$}}
\newcommand{\Vevapoiii}{{\bf V$\rm_{evap,[O~III]}$}}
\newcommand{\VevapHii}{{\bf V$\rm_{evap,H~II}$}} 
\newcommand{\Vco}{{\bf V$\rm_{CO}$}}
\newcommand{\mG}{$\mu$G}
\newcommand{\Smif}{{\bf S$\rm _{mif}$}}
\newcommand{\Smifnii}{{\bf S$\rm _{mif,[N~II]}$}}
\newcommand{\Smifoiii}{{\bf S$\rm _{mif,[O~III]}$}}

\newcommand{\Slow}{{\bf S$\rm _{low}$}}
\newcommand{\Slownii}{{\bf S$\rm _{low,[N~II]}$}}
\newcommand{\Slowoiii}{{\bf S$\rm _{low,[O~III]}$}}
\newcommand{\Sscat}{{\bf S$\rm _{scat}$}}

\begin{document}

\title{Deciphering the 3-D Orion Nebula-I: Expanding Shells in the Huygens Region}


\author{C. R. O'Dell\affil{1}}
\affil{Department of Physics and Astronomy, Vanderbilt University, Nashville, TN 37235-1807}
\author{N. P. Abel\affil{2}}
\affil{MCGP Department, University of Cincinnati, Clermont College, Batavia, OH, 45103}
\and

\author{G. J. Ferland\affil{3}}
\affil{Department of Physics and Astronomy, University of Kentucky, Lexington, KY 40506}

\begin{abstract}
Based on imaging and spectroscopic data, we develop a 3-D model for the Huygens Region of the Orion Nebula. \tC , the hottest star in the Trapezium, is surrounded by a wind-blown Central Bubble that opens SW into the Extended Orion Nebula. Outside of this feature lies a layer of ionized gas at about 0.4 pc from \tC. Both of these features are moving rapidly away from \tC\ with an expansion age for the Central Bubble of only  15,000 yrs. 
\end{abstract}


 \section{Introduction}
 \label{sec:Intro}
\hii\ regions are important in understanding processes such as the abundance of elements throughout our Galaxy. This extends to the study of more distant galaxies because the copious Far Ultraviolet radiation of their hot ionizing stars are efficiently concentrated, through the process of photo-ionization, into easily observed emission lines. What occurs in \hii\ regions also affects the process 
of star formation in massive galactic clusters and may stimulate waves of star formation through their compression of surrounding interstellar gas and dust. The Orion Nebula together with its associated Orion Nebula Cluster is the closest region of star formation that involves massive stars and presents the best opportunity to understand the processes that occur. These may then safely be assumed to operate in more distant and difficult to observe \hii\ regions.

The major goal of the present paper is to use optical data to understand the 3-D structure of the central portion of the Orion Nebula. This will 
include determining the structure from the Main Ionization Front on the surface of the host Orion Molecular Cloud through the outermost shells of atomic gas that cover this region. In a subsequent paper (Paper-II, \citep{ode20b}) we will address how the imbedded Orion-S Cloud affects the structure in the southwest portion of the nebula.

\subsection{Background of this study}
\label{sec:Background}

There is a rich literature on the Orion Nebula \citep{fer01,ode01,mue08,ode08,goi15,kong,pabst}. Most of the emission occurs in 
an ionized blister of gas on the concave Photon Dominated Region (PDR) within the facing surface of the host Orion Molecular 
Cloud. As the gas flows towards the observer from the PDR it is ionized and accelerated. In the vicinity of the strong stellar wind 
from \tC\ one expects a hot bubble to be created, which was identified and characterized from Hubble Space Telescope (henceforth HST) images \citep{ode09}.  Proceeding further towards the observer one encounters a layer of ionized gas 
called Component I in
\citet{abel19} and we designate as the Nearer Ionized Layer (henceforth the NIL) and then two layers of atomic gas (together called the Veil) \citep{vdw89,vdw90,abel04,abel06,vdw13,abel16}. 

The brightest part of the nebula is called the 
Huygens Region (after the first astronomer to publish a drawing of the Orion Nebula). With a characteristic angular size of 4\arcmin\ (0.4 pc),\footnote{We adopt the distance of 383$\pm$3 pc derived by \citet{mk17}, which is in agreement with more recent results using Gaia DR2 \citep{gro19}. This distance gives a scale of 1.86$\times$10$^{-3}$ pc/\arcsec.}
 it occupies the northeast corner of a much larger structure (34\farcm7\ (3.9 pc) north-south and  30\farcm1\ (3.3 pc) east-west). The portions outside of the Huygens Region are 
called the Extended Orion Nebula (EON)\citep{gud08}. The inner atomic component of the Veil appears to be a shell of material that envelopes the entire nebula (both the Huygens Region and the EON) \citet{pabst} with a line of sight (LOS) separation from \tC\ of about 2.0 pc \citet{abel19}. 

 The dominant ionizing star in the Huygens Region is \tC , lying about 0.15$\pm$0.05 pc \citep{ode17} in front of the Main Ionization Front (MIF), with the next most important star being \tA\ \citep{ode17} that lies 135\arcsec\ to the southeast. Along the LOS towards the MIF and its underlying PDR, the Spectral Energy Distribution (SED) of \tC\ 
is modified as it passes through the ambient gas, first in a \ozone\ region that hosts the \oiii\ (500.7 nm) emission and then through a thin \nzone\ layer that produces the
\nii\ (658.3 nm) emission close to the MIF. An underlying consideration in the analysis of the MIF emission is that the expected thickness of the \ozone\  zone that produces the \Vmifnii\ emission should be thinner than the \nzone\ zone that produces the \Vmifoiii\ emission. \citet{bal91} estimated that the \Hplus\ emitting zone (which is where most of the \oiii\ emission occurs) is about 0.09 pc, corresponding to 48\arcsec\ at our adopted distance. \citet{ode18} cites model predictions for e$^{-1}$ thicknesses of 0.0012 pc (0\farcs6) for \nii\ and 
0.026 pc (14\arcsec) for \oiii. In the same paper a profile of the Bright Bar (that must be tilted close to the LOS) gives thicknesses of 2\farcs7 (0.005 pc) and
8\arcsec\ (0.015 pc) for \nii\ and \oiii\ respectively, with the former value probably being an upper limit since the Bright Bar was not resolved below that size.

The 3-D structure of the Huygens Region has been the subject of multiple studies \citep{wen93,ode09,ode17}. These can be summarized as a concave structure marked by two significant features. To the southeast of \tC\ is a linear escarpment (seen as a bright, low ionization, linear feature called the Bright Bar crossing most of the Huygens Region) and a bump in the surface to the southwest of \tC\ commonly associated with the  Orion-S Cloud.

There are three regions of recent star formation within the Huygens Region. The only one visible in the optical is the eponymous Orion Nebula Cluster. 60\arcsec\ 
at position angle 336\arcdeg\ from \tC\ is the highly imbedded BN-KL young star region visible only in infrared and radio wavelengths except for shocks at the tips of fingers radially distributed about a common 
center. The  motion of the optical shocks give an upper limit for their age of 1000 yrs \citep{doi04} and their origin is most likely to be a dynamical event 
that arose some 500 yrs ago \citep{rod05,gomez05}. Knowledge of the proper motion, radial velocity, and direction of the origin place it 0.2 pc behind the local MIF \citep{doi04}.  
The third star formation region lies to the southwest of the Trapezium stars and is the subject of Paper-II. 

A less obvious feature that is important to this study is an arcuate structure structure surrounding \tC . Its reality was established in \citet{ode09} where it was shown to be composed of 
three arcs of \nii\ and \oiii\ emission (designated there as the \oiii\ Shell, the Big Arc east, and Big Arc South and in this study collectively as the High Ionization Arc). This feature is approximately circular near the Trapezium stars, but opens to the southwest, in the direction of the Orion-S Cloud. \citet{gar08} establish that it has a characteristic radial velocity of 10 \kms\ and extends as far east as the Right Ascension of \tA.

\subsection{Nomenclature}
\label{sec:nomenclature}
A note on the nomenclature of this paper is in order. 
Large Samples are areas of 10\arcsec $\times$10\arcsec\ within which spectra from a spatially resolved atlas of spectra of certain emission lines have been averaged. 
Regions are groupings of Large Samples.
The Huygens Region is the brightest part of the Orion Nebula and is in the northeast corner 
of the Extended Orion Nebula. 

Velocities are always given in \kms\ in the Heliocentric reference frame and can be converted to the LSR velocity by subtracting 18.1 \kms.

Directions such as Northeast and Southwest are often expressed in short form as NE and SW.

\subsection{Outline of this paper}
\label{sec:outline}

After presenting the background to the subject of this paper (Section~\ref{sec:Background}) and the outline presented in this section, we present the observational data that we use, their sources, and how we extracted the information used in this paper (Section~\ref{sec:obs}). Testing and determination of the photo-evaporation model is presented in Section~\ref{sec:EvapTest}.  The characteristics and origin of the weaker velocity components are the subject of Section~\ref{sec:Origins}. The regions of locally high extinction are evaluated in Section~\ref{sec:ClearDark}. All of the observations are used to develop a 3-D model of the nebula in a LOS towards the Trapezium, including a calculation of photo-ionization models for the NIL are given 
in Section~\ref{sec:3D}. The properties of the Central Bubble, the colliding layers, a putative relation between the \Vlow\ and \Vmif\ components, and a recently presented alternative 3-D model are discussed in Section~\ref{sec:discussion}. Our conclusions are summarized in Section~\ref{sec:conclusions}.  In the Appendices we illustrate how the observed velocity values are divided into components, how the visibility of weak components on the shoulders of 
strong components depends on the Full Width at Half Maximum  of the strong component, how the magnetic field of the PDR varies in the vicinity of the Orion-S Cloud, and a revised table of velocities in the central Huygens Region is given.

\section{Observations} \label{sec:obs}
\label{sec:observations}

As in our earlier studies \citep{ode18,abel19} we have drawn on the high spectral resolution Spectroscopic Atlas of Orion Spectra \citep{gar08} (henceforth `the Atlas'). The Atlas was compiled from a series of north-south spectra at intervals of 2\arcsec\ and have a velocity resolution of 10 \kms.
The resolution along each slit was seeing limited at about 2\arcsec. 

We employed emission line images made with the Hubble Space Telescope \citep{ode96,ode09} that isolate diagnostically useful emission lines 
covering the Huygens Region. 
    
We use the results from \cite{goi15}. That study included all of the Huygens Region but did not go extensively into the EON. It reported on Herschel satellite spectra of the 158 $\mu$ \Cii\ line at 0.4 \kms\ and 11\farcs4\ resolution. The study also presented H41$\alpha$ observations with the IRAM-30~m telescope. These had 0.65 \kms\ and 27\arcsec\ resolution. Their discussion also used CO 2--1 observations by \citet{ber14} at 0.4 \kms\ and 11\arcsec\ resolution.
        
\subsection{Large Samples of Spectra}
\label{sec:ObsLargeSamples}
\begin{figure}
  \includegraphics
	[width=\columnwidth]
 {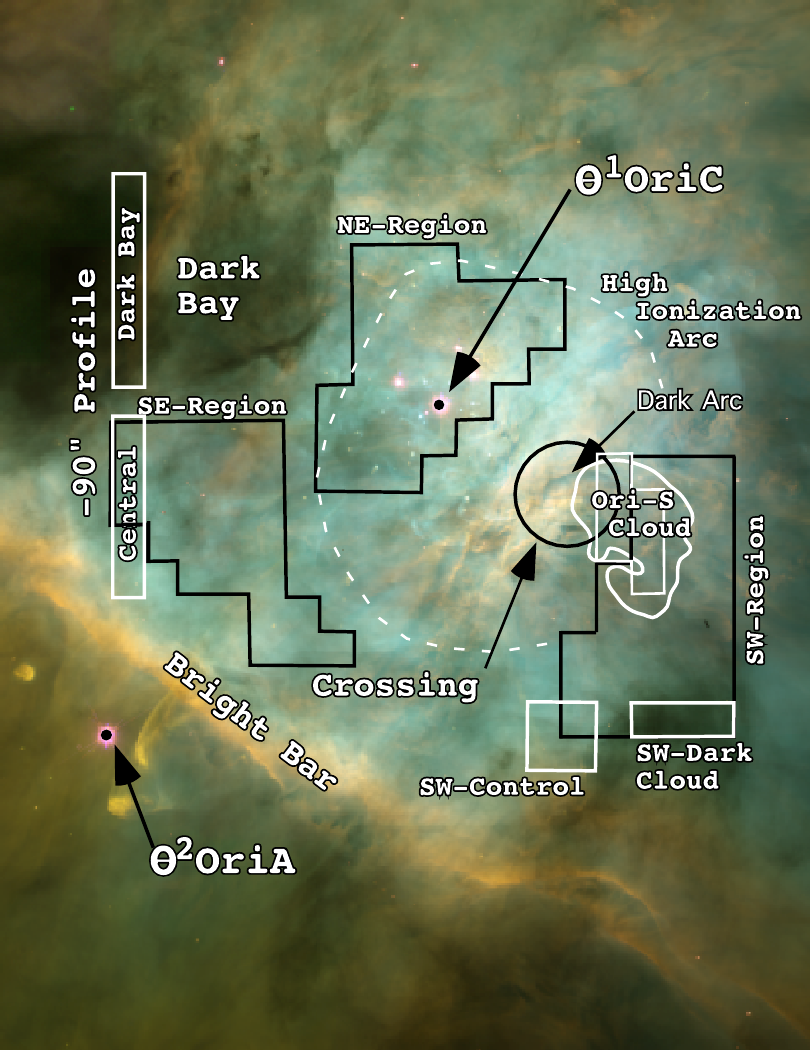}	
    \caption{This 233\arcsec $\times$302 \arcsec\ (0.43$\times$0.56 pc) image centered 36\arcsec\ at PA=166\arcdeg\ from \tC\ is a portion of the Huygens Region \citep{ode96} and is color coded: Blue  \oiii , green \Ha , and red  \nii\ emission. North is up, west is to the right. North is up and East is left. The white solid curved line indicates the 
    edge of the Orion-S Cloud as seen in \hi\ 21 cm absorption \citep{vdw13}. The dashed white line indicates the position of the High Ionization Arc. The NE-Region studied in \citet{abel19} plus the SW-Region and SE-Regions added in the current study are shown with
 black boundaries.The white boxes indicate samples taken to study high extinction regions. The black circle labeled 'Crossing' indicates the region thought to be the most important area for studying the optical features at the NE boundary of the Orion-S Cloud. It is most completely designated as the Ori-S or Orion-S Crossing, but frequently called here the "Crossing".  
 }
\label{fig:fig1}
\end{figure}

\citet{ode18} grouped spectra from the Atlas into averages over areas of 10\arcsec$\times$10\arcsec\ designated here as Large Samples. The higher signal to noise 
(S/N) ratio of these Large Samples were at the expense of spatial resolution. \citet{ode18} evaluated 65 Large Samples, collectively calling them the NE-Region. 
\citet{abel19} used 32 of these Large Samples to define an area also called the NE-Region, a name we use in the present paper because we build upon the \citet{abel19} paper.
The NE-Region was used in \citet{abel19} to study a large column towards \tC\ that was expected to be free of the effects of the Orion-S Cloud.
In order to characterize conditions in a broader area we have employed a grouping of 32 samples called the SW-Region, and another of 27 Large Samples designated as the SE-Region. In addition, samples composed of multiple Large Samples used to study regions of high extinction in \citet{abel19} were used. 
All of these large regions are shown in Figure~\ref{fig:fig1}. 

\subsection{Characteristic Velocity Systems}
\label{sec:VelSys}

The use of the IRAF\footnote{IRAF is distributed by the National Optical Astronomy Observatories, which is operated by the Association of Universities for Research in Astronomy, Inc.\ under cooperative agreement with the National Science foundation.}
 task `splot' has shown that deconvolution of the Atlas spectra \citep{ode18} and earlier higher velocity resolution spectra \citep{hoc88,ode92a,wen93,doi04} reveal common emission line features in multiple areas. For strong lines the accuracy of the derived velocity is about 1 \kms. In this paper we often give velocities to 0.1 \kms (most meaningful when averaging large numbers of velocities) but round-off to the nearest integer when the uncertainty is large. 
 
 We have used the distinguishing properties adopted in \citet{abel19} (only slightly different from those used by \citet{ode18}) for different velocity systems. \Vblue\ components $\leq$-10 \kms\ are assumed to belong to outflows from young stars that create shocks in the ambient nebular gas \citep{ode08}, rather than the large scale pattern of the \Vlow\ spectra. They are not used in our analysis.

 In Appendix~\ref{sec:FWHM} we present a critical analysis of the accuracy of measuring a weak
line on the short wavelength shoulder of a strong emission line. There we see that the accuracy is largely determined by the Full Width at Half Maximum (FWHM) of the strongest (\Vmif) component. 

\subsubsection{Deconvolution of Large Samples}
\label{sec:obsLgSamples}

\begin{figure}
\includegraphics
[width=\columnwidth]
{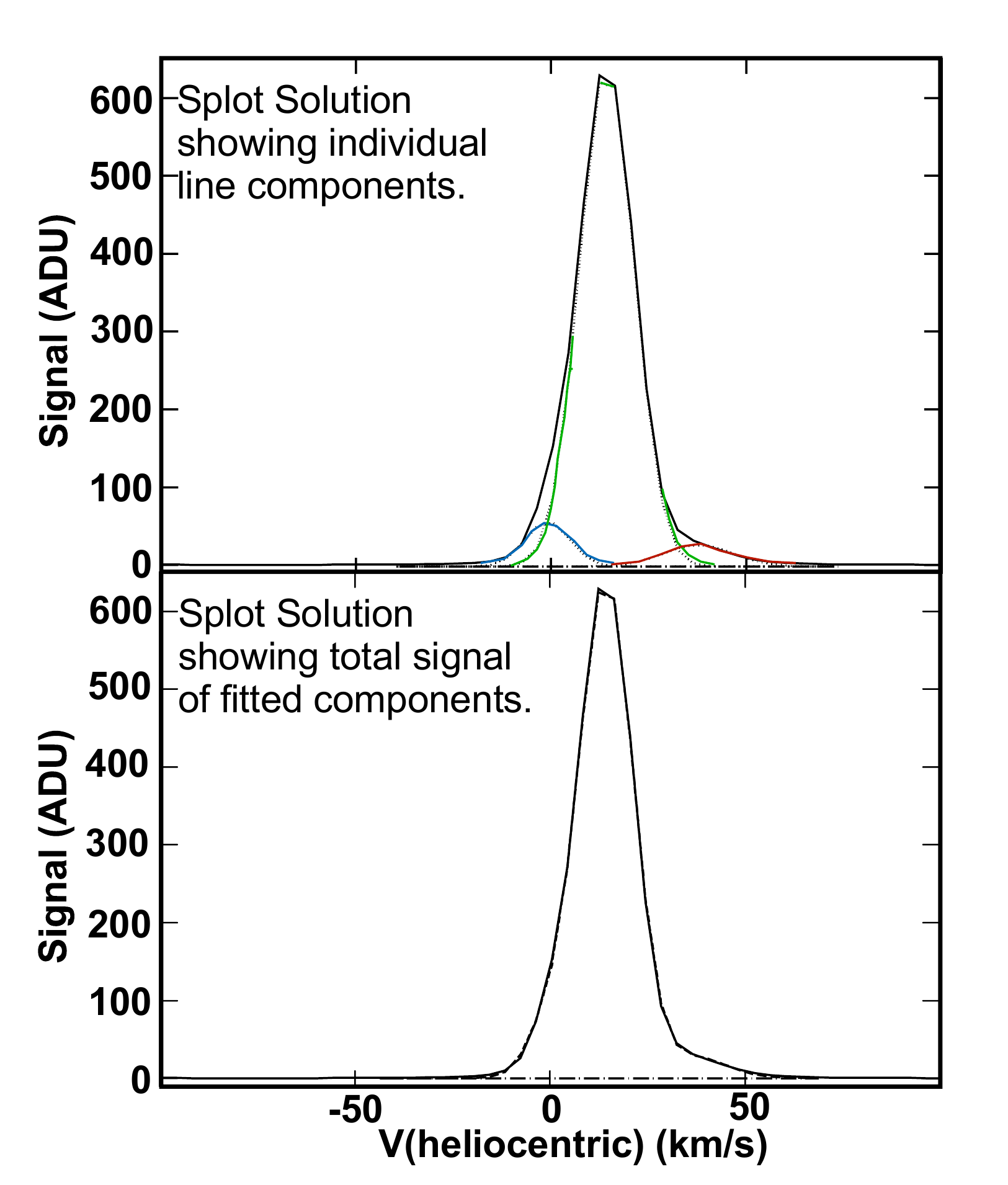}
\caption{The results of using package "splot" to fit the observed line profile of the \oiii\ 500.7 nm line is shown.
The lower panel shows that the fitted accumulative components are indistinguishable from the observed profile. The upper panel shows the individual components whose sum produces the fit in the lower panel.
}
\label{fig:Splot}
\end{figure}

\begin{figure}
 \includegraphics
[width=\columnwidth]
 {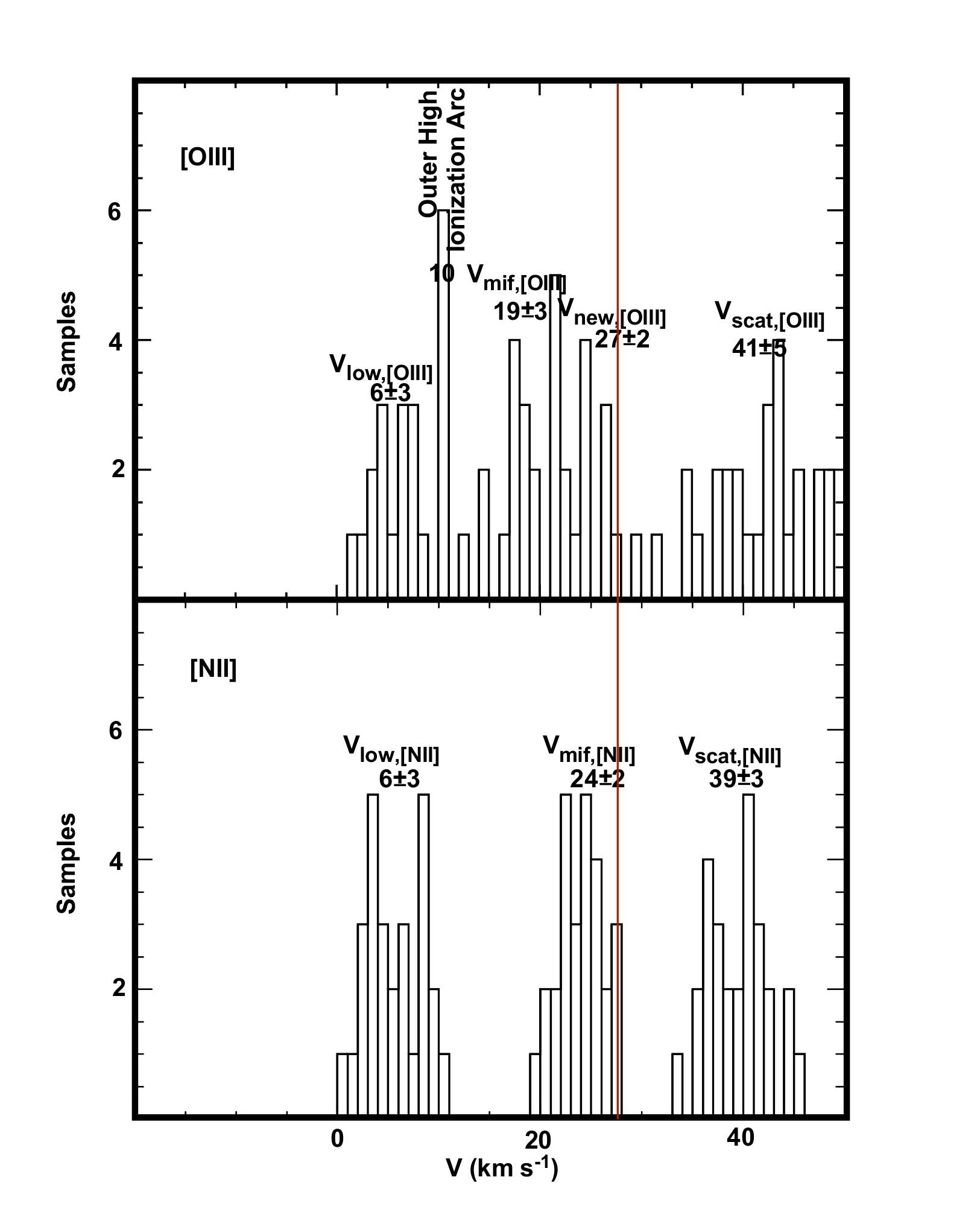}
\caption{These histograms for the SE-Region show the frequency of line components in both \nii\ and \oiii\ and are discussed in Section~\ref{sec:obsLgSamples}. The red line indicates the adopted velocity of the PDR (27.3 \kms ).}
\label{fig:fig3}
\end{figure}

An illustrative sample of the deconvolution of a Large Sample is shown in Figure~\ref{fig:Splot}. In this case the velocity separations and relative strengths are: MIF 0 \kms ,1.00; \Vlowoiii\ 15.4 \kms ,0.07; \Vredoiii\ +23.3 \kms ,0.05. 
 The accuracy of a very weak component lying on the shoulder of the strong \Vmif\ component is discussed in Appendix~\ref{sec:FWHM}.  The results of the deconvolution of the NE and SW-Regions have been published in \citet{abel19} and in Figure~\ref{fig:fig3} we show a histogram of velocity components in the SE-Region, together with their average velocities. The peak in occurrence in \Voiii\ at 10 \kms\ is due to the  High Ionization Arc, as shown in Figures 2 and 15 of \citet{doi04}, where it should be noted that he expresses relative velocities with respect to 18 \kms.  No \Vnewnii\ components are seen in the lower panel, while the upper panel shows the difficulty in distinguishing between \Vmifoiii\ and \Vnewoiii\ components. The relative strength of components helps to distinguish these velocity-ambiguous components.
The results for the deconvolution of the Regions are given in Table~\ref{tab:Regions}.

\begin{table*}
\caption{Results for Regions}
\label{tab:Regions}
\begin{tabular}{lcccccc}
\hline
\hline
 ~&NE-Region&NE-Region&SE-Region&SE-Region&SW-Region&SW-Region\\
\hline
\colhead{Component}&
\colhead{\nii}&
\colhead{\oiii}&
\colhead{\nii}&
\colhead{\oiii}&
\colhead{\nii}&
\colhead{\oiii}\\
\hline
\Vscat  &40.3$\pm$3.4 & 37.9$\pm$3.1&39.4$\pm$3.1&41.0$\pm$5.2&38.8$\pm$3.2& 37.6$\pm$2.6\\
\Vnew   &36.5$\pm$1.8 & 27.4$\pm$5.4 & --- &26.5$\pm$1.7& 33.3$\pm$2.2 & 26.9$\pm$7.6\\
\Vmif     & 22.4$\pm$2.2&18.0$\pm$2.8 & 23.8$\pm$2.3&19.0$\pm$2.9 &18.0$\pm$1.8 & 10.8$\pm$2.8\\
\Vlow    & 5.6$\pm$3.7 & 7.8$\pm$2.1 & 5.5$\pm$2.7&5.5$\pm$3.0& 2.8$\pm$1.9 & 0.7$\pm$3.3\\
\Vblue   & -6.1$\pm$2.8&0.6$\pm$3.8& -12.3$\pm$1.5**& -4.8***& -4.2$\pm$3.1&-0.8$\pm$4.7\\
\Sscat /\Smif\ & 0.06$\pm$0.03 & 0.06$\pm$0.03&0.060$\pm$0.026 &0.048$\pm$0.034 & 0.06$\pm$0.03 &0.07$\pm$0.03\\
\Slow /\Smif\ & 0.10$\pm$0.03 & 0.13$\pm$0.12 &0.25$\pm$0.11 & 0.13$\pm$0.08---0.95$\pm$0.26*&0.10$\pm$0.03 &0.09$\pm$0.02---0.65$\pm$0.19*\\
\hline
\end{tabular}\\
\newline
~* Values cluster around these two values.

** Two Large Samples only.

*** One Large Sample only.
\end{table*}

 \section{Testing the photo-evaporation model}
\label{sec:EvapTest}

The basic model of the Huygens Region being a thin ionized layer on the surface of the host Orion Molecular Cloud (OMC) was recognized from the expectation that a photo-evaporating cloud would accelerate as it left the PDR with observations that agreed with the theoretical expectation. The original papers independently proposing this model \citep{zuk73,bal74} drew on the limited velocity information then available, recognizing that the higher ionization material (further from the PDR and closer to \tC) had a more negative radial velocity than the lower ionization material. This model has been refined by recognition of the concave structure of the MIF \citep{wen95} and recognition that the velocity variations across the nebula but within the same ion are due to local smaller scale variations in the tilt of the PDR \citep{ode18}. 

The \Vmif\ values in Table~\ref{tab:Regions} are all lower than the reference \Vpdr\ velocity of 27.3 $\pm$0.3 \kms\ that we have adopted \citep{ode18}. \Vmifnii\ values always higher than \Vmifoiii, consistent with the \nzone\ layer (producing \Vmifnii\ emission) not having been accelerated away from the PDR as the \ozone\ layer that produces the \Vmifoiii\ emission.

The \Vevap\ values derived from our assumed \Vpdr\ and the \Vmif\ NE-Region and SE-Region values of Table~\ref{tab:Regions} are \Vevapnii\ = 4$\pm$3 \kms\ and \Vevapoiii\ = 9$\pm$3 \kms. Quite different values are derived from the SW-Region (\Vevapnii\ = 9$\pm$2 \kms\ and \Vevapoiii\ = 16$\pm$3 \kms). The differences in the derived \Vevap\ for the regions could be that the SW-Region is more nearly flat-on as viewed by the observer than the other two regions. 

The photo-evaporation model can also be testing using the results of \citet{goi15}, who studied 158 $\mu$ \Cii\  emission
across the nebula, with particular emphasis on a region centered on \tC , which lies within our NE-Region. 
This radiation arises within the PDR. Immediately behind the PDR lies the CO emitting layer in the background molecular cloud. Their Fig.~3 shows that \Vcii\ and \Vco\ are at 28 \kms\ (10 \kms\ in the LSR velocity system they use). Recombination hydrogen emission H41$\alpha$ is at 16 \kms\ (-2 \kms\ LSR). Since ionization models show that the hydrogen radiation should arise mostly from the \oiii\ emitting layer, these data indicate that \VevapHii\ = 12$\pm$2 \kms , greater than \Vevapoiii\ = 9$\pm$3\kms and  8$\pm$3 for the NE and  SE Regions,
but less than 16$\pm$3 for the SW-Region. 

For the remainder of this paper we will adopt \Vevapnii\ = 7$\pm$4 \kms\ and \Vevapoiii\ = 12$\pm$4 \kms.

\section{Origin of the weaker velocity components seen in the Large Samples}
\label{sec:Origins}

The strongest velocity component (\Vmif) arises from the MIF lying immediately on the observer's side of the ionization front along the PDR. In this section we will illustrate the features and the interpretation of the weaker components within the Large Samples. A summary table of all the velocities except those in the two outermost Veil components is give in Appendix~\ref{app:revisions}.

\subsection{Origin of the \Vscat\ Component}
\label{sec:Vscat}

A red-shifted component is found in almost all of the Large Samples and the Profile Samples. After discovery that the nebula's continuum was strong \citep{green39a,green39b}, the sources that dominate in the continuum (the Trapezium stars) was established quantitatively by \citet{bal91}.
In their comprehensive lower resolution study, \cite{ode10} demonstrated that scattering of nebular emission occurs not only in backscattering by the nearby PDR but also at large distances. However, the red-shifted component arises from local backscattering.

The \Vscat\ components in the Huygens Region are usually attributed to backscattering by dust in the PDR \citep{ode92b,hen98,ode01,abel06,ode18} lying along the same LOS. Within this model where the emission lines arise from photo-evaporation of ionized material away from the ionization front on the PDR, the expected red-shift of the backscattered component, relative to the 
layer producing the \Vmif\ emission, is about twice the photo-evaporation velocity \citep{hen98}. Since \Vevapoiii\ is larger than \Vevapnii , the \Vscatoiii\ has a larger red-shift and it is easier to detect. The observed and predicted values of \Vscat\ - \Vmif\ are given in Table~\ref{tab:DeltaVscat}.
Where we see that \Vscatnii\ - \Vmifnii] values are somewhat larger than expected from our adopted \Vevapnii\ and \Vscatoiii\ - \Vmifoiii\ are in good agreement.

\begin{table}
\caption{Separation of \Vscat\ and \Vmif\ Components}
\label{tab:DeltaVscat}
\begin{tabular}{lcc}
\hline
\hline

\colhead{Region}&\colhead{\Vscatnii\ - \Vmifnii}&\colhead{\Vscatoiii\ - \Vmifoiii }\\
\hline 
NE-Region& 18$\pm$4&19$\pm$4\\
SE-Region& 16$\pm$3&22$\pm$6\\
SW-Region& 21$\pm$3&27$\pm$3\\
Average&  18$\pm$4 & 23$\pm$5\\
Predicted   & 14$\pm$4 & 24$\pm$4\\
\hline
\end{tabular}\\
\end{table}

\begin{table}
\caption{\Sscat /\Smif}
\label{tab:ScatOverSmif}
\begin{tabular}{ccc}
\hline
\hline 
\colhead{Region}&\colhead{\nii }&\colhead{\oiii}\\
\hline 
NE-Region& 0.05$\pm$0.02&0.06$\pm$0.026\\
SE-Region& 0.07$\pm$0.06&0.04$\pm$0.027\\
SW-Region& 0.07$\pm$0.03&0.08$\pm$0.024\\
Average&  0.06$\pm$0.04 & 0.06$\pm$0.03\\
\hline
\end{tabular}\\
\end{table}

Table~\ref{tab:ScatOverSmif} shows that the signal of the \Vscat\ component  relative to the \Vmif\ component varies little between the Regions and is indistinguishably the same in \nii\ and \oiii.

Models of artificial spectra using the methods of Appendix~\ref{sec:FWHM}
 and the average FWHM of 16.4 \kms\ and 13.2 \kms\ for \nii\ and \oiii\ showed that the derived properties of the scattering components (\Vscat\ and \Sscat /\Smif\ agree well with the predictions of models. That is to say that the wide separation of the \Vmif\ and \Vscat\ components allow  derivations unaffected by blending.

\subsection{Origin of the \Vlow\ Component}
\label{sec:Vlow}

The component that we call \Vlow\ is associated with the NIL (called \Vlow\ and Ionized Component I in \citep{ode18} and \citep{abel19} respectively). 
Clear evidence for it in the direction of the Trapezium and \tA\ lie in absorption lines formed there. These are \heI\ at 2.1$\pm$0.6 \kms\ \citep{ode93}, \Piii\ at 4.9$\pm$3.0 \kms\ \citep{abel06}, \siiia\ at 4.5$\pm$0.9 \kms\ \citep{abel06}, \Caii\  at 7.5 \kms\ \citep{ode93}, and \Nai\ at 6.0 \kms\ \citep{ode93}, for an average of 5.0$\pm$2.0 \kms. 

Accurate emission line velocities for the \Vlow\ components are more difficult to determine because the lines fall on the shoulder of the much
stronger \Vmif\ component. In Appendix~\ref{sec:FWHM} we demonstrate that the limit of detectability of the separation from \Vmif\ is determined by the FWHM of the \Vmif\ component, this lying about 0.5 \kms\ below the FWHM.

 For \nii\ the average FWHM is 16.4$\pm$0.6 \kms\ and the average
splitting \Vmifnii\ - \Vlownii\ in Table~\ref{tab:Regions} is 17$\pm$4, thus the average \Vlownii\ = 5$\pm$3 indicates that the \Vlownii\ component arises from the same layer as the absorption lines (the NIL).  

The FWHM for the \Vmifoiii\ component is 13.2$\pm$0.8 \kms\ and the average  
splitting \Vmifoiii\ - \Vlowoiii\ = 11$\pm$3, indicating that \Vlowoiii\ lines are affected by the difficulty of extracting the \Vlowoiii\ from the shoulder of the \Vmifoiii\ component.  In the much higher resolution study of \oiii\ by \citet{hoc88} (FWHM = 10.9$\pm$1.9 \kms) \Vmifoiii\ - \Vlowoiii\ = 8.0$\pm$2.6 \kms, which is evidence that that the average \Vlowoiii\ in the Regions is 8$\pm$3 \kms. Caste\~neda  employed the KPNO Coude Spectrograph with a resolution $\lambda$/$\delta \lambda$ of 100,000, giving an instrumental FWHM of 4.0 \kms\ which clearly identified the \Vmifoiii\ FWHM. \Vlowoiii\ derived from our value of \Vmifoiii\ and his separation (\Vlow\ = 8$\pm$3 \kms\ is adopted in the remainder of this report.
 
This value again indicates that the \Vlowoiii\ component arises from the same layer as the NIL.

Two additional studies using the same velocity resolution as the \citet{hoc88} study give \Vlow\ velocities of 3 \kms for \oii\ \citep{jones92} and 10 \kms for \siii\ \citep{wen93}, again indicating association with the NIL. 

\begin{figure}
 \includegraphics
[width=\columnwidth]
 {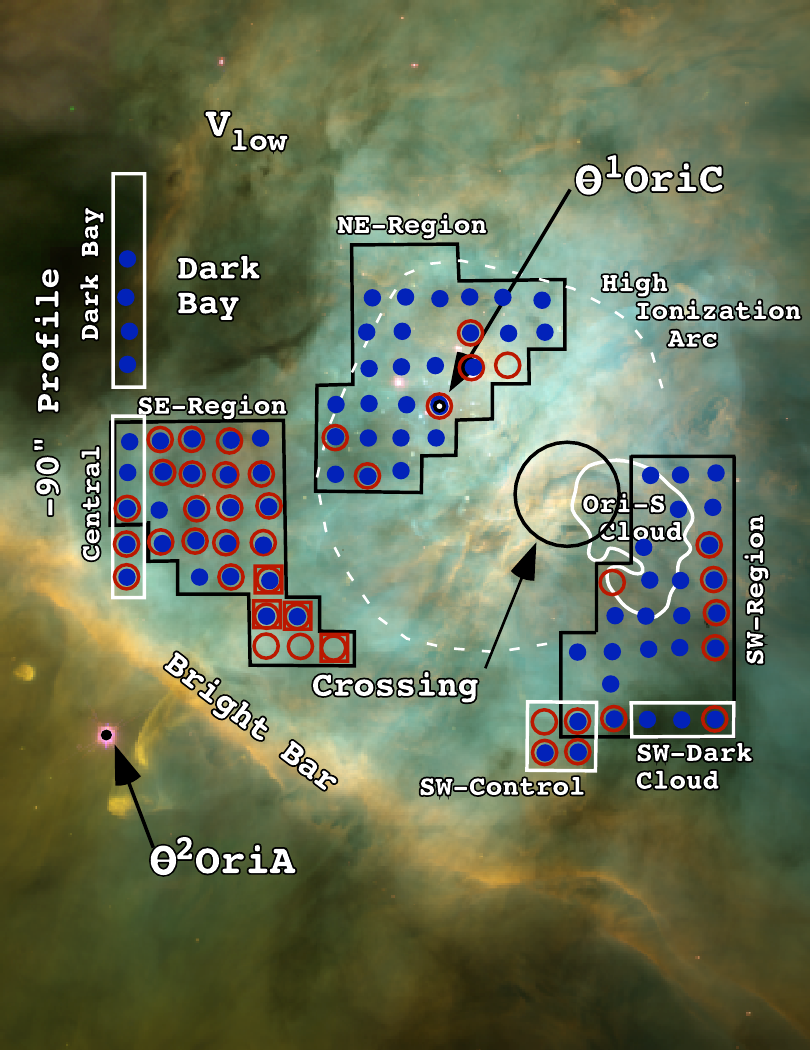}
\caption{Like Figure~\ref{fig:fig1} except now we indicate where a \Vlow\ component is detected (blue filled circles \Vlownii, red circles \Vlowoiii ). The large open squares indicate where the \Slowoiii /\Smifoiii\ ratio is unusually large.) 
}
\label{fig:fig4}
\end{figure}

The distribution of where the \Vlow\ components are seen are shown in Figure~\ref{fig:fig4}.
The distribution may extend as far as the LOS towards \tA. High resolution spectroscopic study of this star \citep{ode93} shows  
two \heI* absorption components at -2.9 \kms\ and 5.2 \kms, (with uncertainties of about 1 \kms ). The more positive can be associated with the average of the Regions  \Vlowoiii\ = 5$\pm$3 \kms\ and would be evidence that the NIL 
system extends beyond the inner Huygens Region. Because the 388.9 nm absorption line is formed in gas of the same level of ionization as the \Vlowoiii\ emission line, the conclusion of the NIL extending as far as \tA\ is strengthened by the fact that \Vlowoiii\ for the nearby SE-Region is 6$\pm$3 \kms.

We can safely conclude that the NIL is real, extends across the inner Huygens Region, has a velocity of about 6$\pm$2 \kms , and is most visible in \nii.

\subsection{Origin of the \Vnew\ and \Vblue\ Components}
\label{sec:newblue}

\begin{figure*}
 \includegraphics
[width=7in]	
 {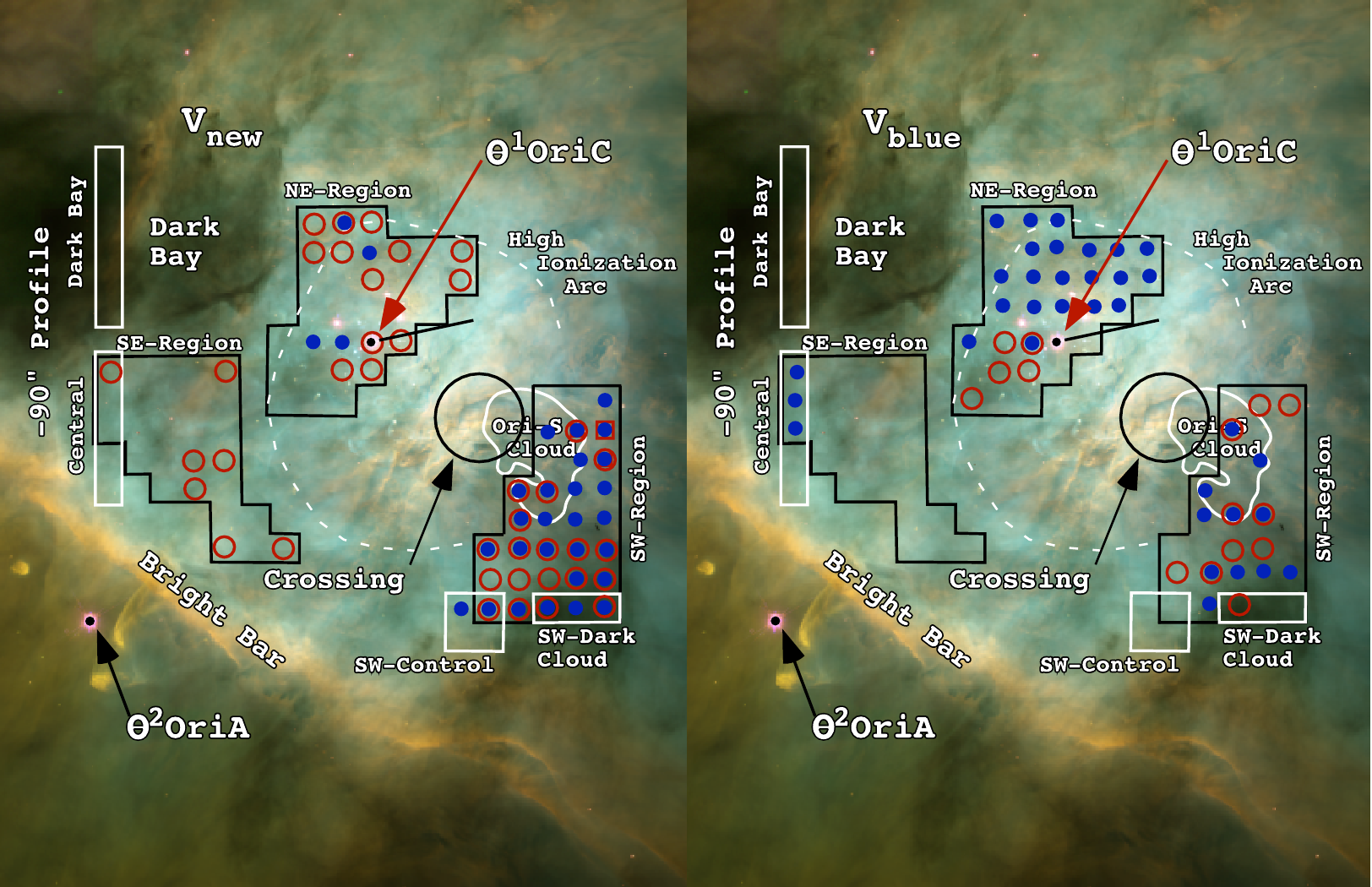}
\caption{Like Figure~\ref{fig:fig4} except now we indicate where the 
\Vnew\ (left panel) and \Vblue\ (right panel)  components are detected (blue filled circles indicate \nii\ detections and red circles \oiii\ detections.) 
}
\label{fig:fig5}
\end{figure*}

The {\bf \Vnew} components were originally reported in \citet{ode18}, where they were only detected in \oiii, while \citet{abel19} found it in \nii. 
Our study includes many more large samples than in these previous studies and the location of the detected \Vnew\ components are shown
in the left-hand panel of Figure~\ref{fig:fig5}. The average \Vnewnii\ in the two regions were it is detected is 35$\pm$3 \kms , while the \Vnewoiii\ 
average in the three regions is 27$\pm$4. 

The \Vblue\ components are weaker than the \Vlow\ components, even after factoring in the classification criteria. Their locations are shown in 
the right-hand panel of Figure~\ref{fig:fig5}. The average \Vbluenii\ in the two regions where it is frequent is -5$\pm$3 \kms. Similarly 
the average \Vblueoiii\ is 0$\pm$4 \kms. 
Recall that in this tabulation, all of the velocity components $\leq$~-10.0 \kms\ are assumed to belong to outflows from young stars that create shocks in the ambient nebular gas. Including the spectra with \Vlownii\ $\leq$~10 \kms\ would decrease the averages by 2.2 \kms.

\citet{abel19} attributes the \Vblue\ components to material at the approaching side of an expanding hot shocked bubble surrounding \tC , called there the Nearer Central Bubble (we will use Central Bubble). Our larger data-set comes to the same conclusion if the effects of the Central Bubble extend into the regions immediately outside of the High Ionization Arc.  Their attribution of the \Vnew\ components to shocked gas moving into the high density MIL is also acceptable, with \Vnewnii\ (35$\pm$3 \kms) moving  15 \kms faster than the MIF (20$\pm$3). Similarly \Vnewoiii\ (27$\pm$4 \kms) moves 11 \kms\ relative to the MIF (16$\pm$4).

\section{High Extinction Regions}
\label{sec:ClearDark}

The primary source of extinction and reddening in the Huygens Region is the foreground Veil \citep{ode92b,ode00}. 
In the latter study, done at higher spatial resolution, it was shown that the greatest extinction occurs in the region known as the Dark Bay, where the logarithmic \Hb\ extinction coefficient (\cHb) reaches 2.0. This region is crossed by the -90\arcsec\ Profile \citep{ode18} where \cHb\ is about 1.6 and the results are included in Table~\ref{tab:ClearDark} under the heading -90\arcsec\ Dark Bay.
A second region of high extinction is in the direction of the Orion-S Cloud, where \citet{ode00} found \cHb\ about 0.6. This would be a lower limit since the part of the radio continuum used to derive the extinction arises from the side of the Orion-S Cloud that faces the OMC. This region was characterized by 6 Large Samples as shown in Figure~\ref{fig:fig1}. A third region of high extinction 
is the dark SW-Cloud \citep{gar07}, where \cHb\ is about 1.0. Three Large Samples 
were used in \citet{ode18} to isolate the SW-Cloud and another four to isolate a nearby region that appears to be free 
of this extra extinction. The results for these samples are repeated in Table~\ref{tab:ClearDark} under the headings `SW Control' 
and `SW Cloud'. We examine the properties in the three regions and their nearby low extinction areas in order to assess the effects of the extinction and what it can tell us about the structure along the LOS. 

The primary comparison regions for the Dark Bay are the -90\arcsec\ Profile Central sample and the SE-Region. For the Ori-S Cloud, the NE-Region and SW-Regions are useful for comparison. These pairings are in addition to the SW-Cloud and SW-Control Regions. The results for all of the pairings are shown in Table~\ref{tab:ClearDark}. 

In the Dark Bay, we see a large jump in \Smifnii /\Smifoiii , which is compatible with the high extinction there. 
We probably do not see a \Vlowoiii\ component in the -90\arcsec\ Dark Bay sample because the large FWHM (19.4$\pm$0.6 \kms)of the \Vmifoiii\  component makes it undetectable on the shoulder of the \Vmifoiii\ compnent. 
Four of the six Dark Bay \Vmifnii\ components are very wide (FWHM = 23$\pm$2 \kms). This indicates that it is a blend with another velocity component or that we are seeing a blend of extincted \Vmifnii\ emission and emission arising from the near side of the Dark Bay. 
These same factors could also lead to the jump in \Vmifnii\ above the Control and SE-Region. Similar considerations may explain the jump 
in \Vlownii , although the rise may be an artifact of extracting the \Vlownii\ from the shoulder of the unusually wide \Vmifnii\ component.

In the Orion-S Cloud there is only a marginal indication of a rise in \Smifnii /\Smifoiii , as expected since the extinction is occurring within the Cloud and we see the foreground side of the Cloud. Unfortunately, only one Large Sample record \Vlow components, so no meaningful comparison can be made of these with their surroundings.

In the SW-Cloud \Smifnii /\Smifoiii\ rises to 1.7$\pm$0.1, well above the SW Control region of 1.1$\pm$0.2, and indicating high extinction, although not as great in the Dark Bay. \Slownii /\Smifnii\ is the same as in the nearby SW Control samples and the SW-Region. \Slowoiii /\Smifoiii\ is compatible with one of the unexplained groupings in the SW-Region and in the SW Control samples. 

We note that there is a general decrease in \Vmif\ proceeding from the NE to the SW across of the Huygens Region. This is quite smooth for \Vmifnii , while \Vmifoiii\ drops abruptly at the Orion-S Cloud.
 \Slow /\Smif\ ratios are usually the same (within their probable errors) in the dark samples and their comparison regions. This 
indicates that the layer producing the \Vlow\ components (the NIL) lies between \tC\ and Veil components B and C that cause the extinction.  

\floattable
\begin{deluxetable}{lccccccc}
\setlength{\tabcolsep}{0.02in}
\tabletypesize{\scriptsize}
\tablecaption{Comparison of Large Samples relevant to discussion of local extinction*
\label{tab:ClearDark}}
\tablewidth{0pt}
\tablehead{
\colhead{Sample}&
\colhead{\Vmifnii} &
\colhead{\Vmifoiii}&
\colhead{\Vlownii}&
\colhead{\Vlowoiii}&
\colhead{\Smifnii /\Smifoiii}&
\colhead{\Slownii /\Smifnii}&
\colhead{\Slowoiii /\Smifoiii}}
\startdata
-90\arcsec Central Region &23$\pm$1  &20$\pm$2       &6$\pm$3    &6$\pm$2        &1.7$\pm$1.4      &0.2$\pm$0.1  &0.2$\pm$0.1\\
-90\arcsec Dark Bay          &28$\pm$1  &18$\pm$0.2    &11$\pm$3     &---                  &3.4$\pm$0.6      &0.4$\pm$0.2     &--\\
SE-Region                         &24$\pm$2   & 19$\pm$3      &  6$\pm$3     &6$\pm$3        &1.7$\pm$0.8    &0.25$\pm$0.11   &0.13$\pm$0.08---0.95$\pm$0.26$\dagger$\\
NE-Region                         & 22$\pm$2 &18$\pm$3      & 6$\pm$4      &8$\pm$2         &1.5$\pm$0.5     &0.10$\pm$0.03  &0.13$\pm$0.12\\
Orion-S Cloud                    &21$\pm$1  & 11$\pm$2      &2.6$\dagger\dagger$            &-1.0$\dagger\dagger$           & 1.7$\pm$0.4    &0.08$\dagger\dagger$              &0.07$\dagger\dagger$\\
SW-Region                        &18$\pm$2   & 11$\pm$3     &3$\pm$2      & 1$\pm$3         & 1.6$\pm$0.5    &0.10$\pm$0.03  &0.09$\pm$0.02--- 0.65$\pm$0.19$\dagger$\\
SW Control                        &  17$\pm$2  &11$\pm$3      & -2$\pm$3     & -1$\pm$1     &1.1$\pm$0.2     &0.07$\pm$0.04  &0.07$\pm$0.03\\
SW Cloud                          & 18$\pm$2     &12$\pm$1   &-1$\pm$3       &0$\pm$7        &1.7$\pm$0.1     &0.09$\pm$0.03  &0.06$\pm$0.03\\
\enddata
\tablenotetext{*} {All velocities are Heliocentric and in \kms. LSR values are 18.2 less. Parentheses indicate the reduced number of spectra.}
\tablenotetext{\dagger}{Ratios group around these values.}
\tablenotetext{\dagger\dagger}{One Large Sample only.}
\end{deluxetable}

\section{Development of a 3-D Model for the Line of Sight near the Trapezium}
\label{sec:3D}

Understanding the structure along a LOS from the observer to the PDR requires using all the data in hand, plus computational modeling. Through these we have developed a 3-D model.

\subsection{Refinement of the \tC\ -- NIL photoionization model distance}
\label{sec:refinement}
\begin{figure}
 \includegraphics
[width=3.5in]	
{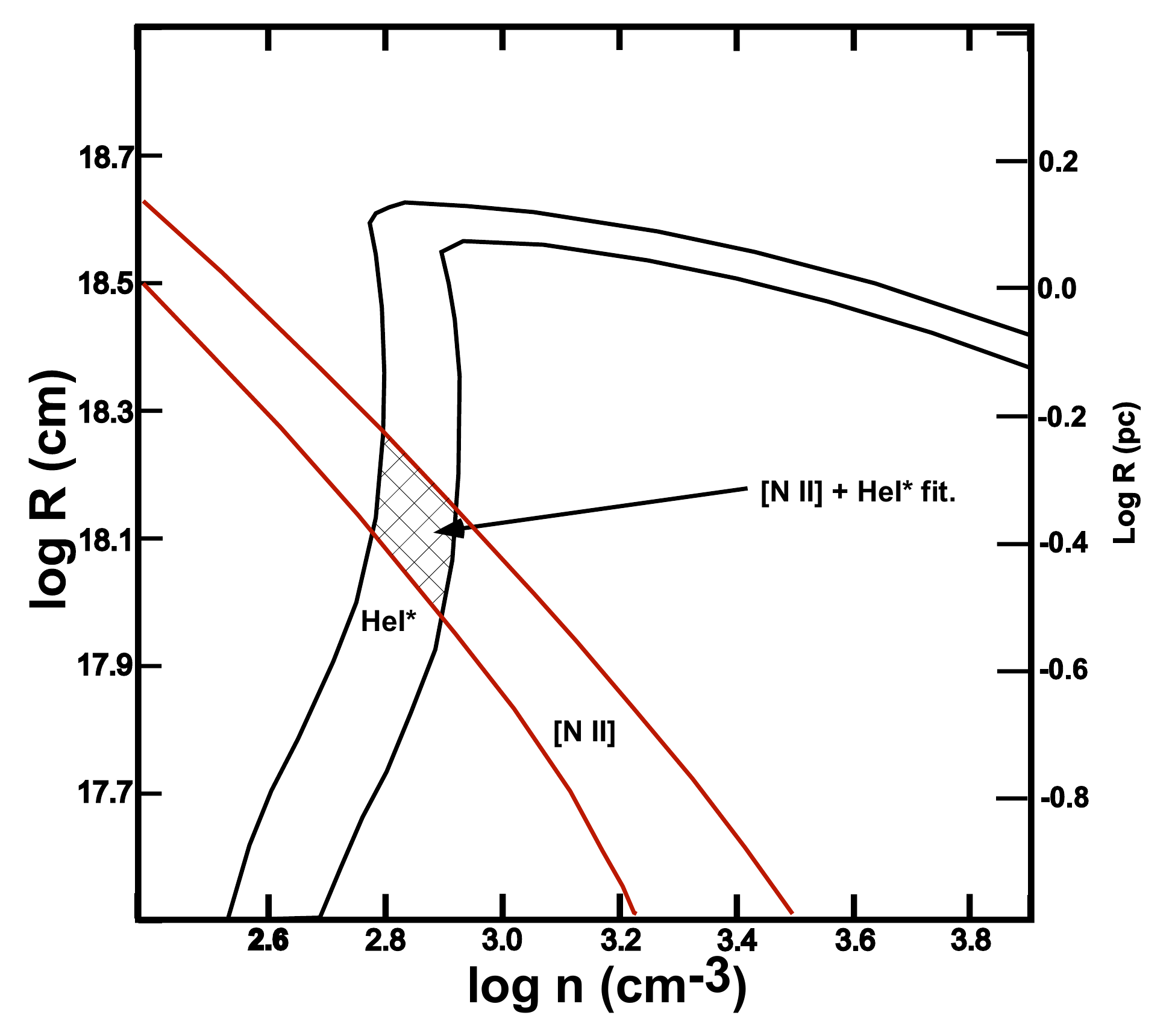}
\caption{This plot of log n (cm$^{-3}$, the density of hydrogen atoms) versus the separation of \tC\ and the NIL is like Figures 4, 5, and 6 in \citet{abel19} except that a larger range of density and distances is used and we have dropped modeling of \oiii.  In addition, improved the accuracy of the \heI*\ and \nii\ observational constraints were utilized. The red lines enclose the range of values where the photoionization models match the observed surface brightness of \Vlownii. The black lines include the range consistent with the observed value of the column density of \heI *. The cross-hatched region indicate where there is agreement of the \nii\ and \heI * data, as discussed in Section \ref{sec:refinement}. 
}
\label{fig:fig6}
\end{figure}

In \citet{abel19} photoionization models were calculated for the NIL over a wide range of distances from \tC\ and the density of the layer. These were then used to predict the surface brightness in the \nii\ and \oiii\   emission lines and the column density of \heI\ in the lowest triplet state(2$\rm ^{3}$S, designated as \heI *), where the 388.9 nm absorption line arises that appears in the spectra of the Trapezium stars. 
We repeated those calculations using an improved determination of the 388.9 nm line and the surface brightness in \nii\, in addition to dropping modeling of the \oiii\ line. 

\citet{abel19} used the observed equivalent width of the 388.9 nm line from the brightest four Trapezium stars to derive an average column density N(\heI*, cm$^{-2}$). We now think it better to use only the observations of \tAt, \tC, and \tD\ since the 388.9 nm absorption 
line in \tB\ is clearly affected by the nearby \heI\  and H8 emission lines. The average of the
three stars is 1.45$\pm$0.17$\times$10$^{13}$cm$^{-2}$ with a better defined uncertainty. 

We have used a higher value of the surface brightness of \nii\ based on artificial models of the emission line, as done in Appendix~\ref{sec:FWHM} except that we have specifically used FWHM (16.6$\pm$1.4 \kms), \Vmifnii\ - \Vlownii\ (15.3$\pm$1.3 \kms), and \Slownii /\Smifnii\ (0.07$\pm$0.02) from
nine Large Samples surrounding \tB\ (the star used to derive the ultraviolet absorption lines). These models established that the 
true \Slow /\Smif\ is 1.4 times larger than derived from 'spcflot'.

We have not modeled the \oiii\ line because \Vlowoiii\ is not usually present, therefore the few available values do not give an accurate 
constraint on the models. As a result, we have one less constraint on the model of the NIL than in \citet{abel19}, but the \nii\ 
 and \heI*\ constraints are now better defined and more reliable.

We see the results of the the comparison of the predictions and observations in Figure~\ref{fig:fig6}. 
The combination of brighter \nii\ emission and a 30$\%$ higher \heI*\ column density yields a model with nearly the same density as in \citet{abel19}, but about half the distance from \tC.
There is a single overlapping zone of allowable fits of the observations and the predictions, shown as a cross-hatched region in Figure~\ref{fig:fig6}. The Central value is Log R = -0.39 (0.41 pc) and Log n = 2.84 (690 \cmq). The allowable range of log R is -0.22 (0.60 pc) to -0.52 (0.30 pc) and the allowable range of Log n is 2.79 (620 \cmq) to 2.91 (790 \cmq).

The 2$\rm ^{3}$S state is populated by recombinations of singly ionized helium and this region also produces the \oiii\ emission, but not \nii\ emission.
This means that a caveat on our conclusions is that the 388.9 nm records the \ozone\ zone and the \nii\ line the \nzone\ zone. Without a detailed model of the ionization structure of the NIL, we cannot quantitatively assess how this could affect our results. However, we expect the effects to be small.

\subsection{Relative Positions of the NIL and the Orion-S Cloud in earlier studies}
\label{sec:discNILandOriS}

The distance between the blue-shifted collimated outflows from the Orion-S Cloud  and material with which it collides has been treated in multiple earlier studies, some predating knowledge of the the NIL. 

In their study of HH~203 and HH~204 that lie to the southeast from the Bright Bar, \citet{doi04} calculated the angles of the flow from the Orion-S Cloud and determined foreground displacements of the shocked material at 0.2 pc and 0.3 pc respectively, coming to the conclusion that these shocks were the result of jets originating from the Orion-S Cloud shocking material in the MIF in the region immediately southeast of the Bright Bar. They accepted this geometry because the NIL was not recognized at that time, the MIF was known to curve towards the observer at the Bright Bar,  and estimates of the Veil atomic components distances were much larger. With the recognition of the NIL, it is more likely  that this is NIL material, shocked by the SE outflows from the Orion-S Cloud.

\citet{vdw13} associated the Veil \hi\ absorption feature F with the well studied flow and shocks defining HH~202 (that lies to the NW from the Orion-S Cloud) and concluded that the Veil lay 0.26 pc (corrected to our distance of 383 pc) towards the observer from the Orion-S Cloud. They did not consider the existence of the NIL. In \citet{abel16} we established that the \Htwo\ absorption spectrum along with the neutral carbon absorption spectrum in the UV makes it impossible for Veil component B to be this distance from the Cloud and therefore the Trapezium. The present work reconciles \citet{abel16} with \citet{vdw13}. There is a dynamical interaction between the layers in front of the Trapezium and HH 202, but the interaction is not with Veil Component B, but with the NIL. 

 In the context of the present study, we note that absorption feature F shows two velocity peaks at -1 \kms\ and +7 \kms\ (both Heliocentric) and that these fall into the range of velocities encountered in the NIL. Linking an \hi\ absorption line area to a small region in the ionized NIL is plausible by assuming mass loading at the front of the flow followed by rapid recombination of ionized hydrogen. This was predicted in the models for Herbig--Haro shocks by \citet{pat87}. \citet{abel16} argue that the association of HH~202 and absorption feature F is incorrect, presenting several arguable if not hard and fast reasons that HH~202 and absorption feature F are not associated. They do not consider that a neutral zone can be formed at the head of a shock.

If one accepts that the 
HH~202, HH~203, and HH~204 shocks occur in the nearest foreground layer, this would mean that the NIL is about 0.2 -- 0.3 pc in front of the Orion-S Cloud, the host of multiple stellar outflows. The distance in front of \tC\ would depend on the separation of \tC\ and the Orion-S Cloud along the LOS.

\subsection{Probable spacings of \tC, the Orion-S Cloud, and the NIL}
\label{sec:FinalDistances}

The position of the features along a LOS near the Trapezium can be determined from multiple lines of evidence. Even though each of these methods has an uncertainty because of assumptions made in each,  together they give a self-consistent explanation of the positions. The model presented is an improvement over previous efforts \citet{ode09,ode10}.

\begin{figure}
 \includegraphics
 [width=\columnwidth]
 {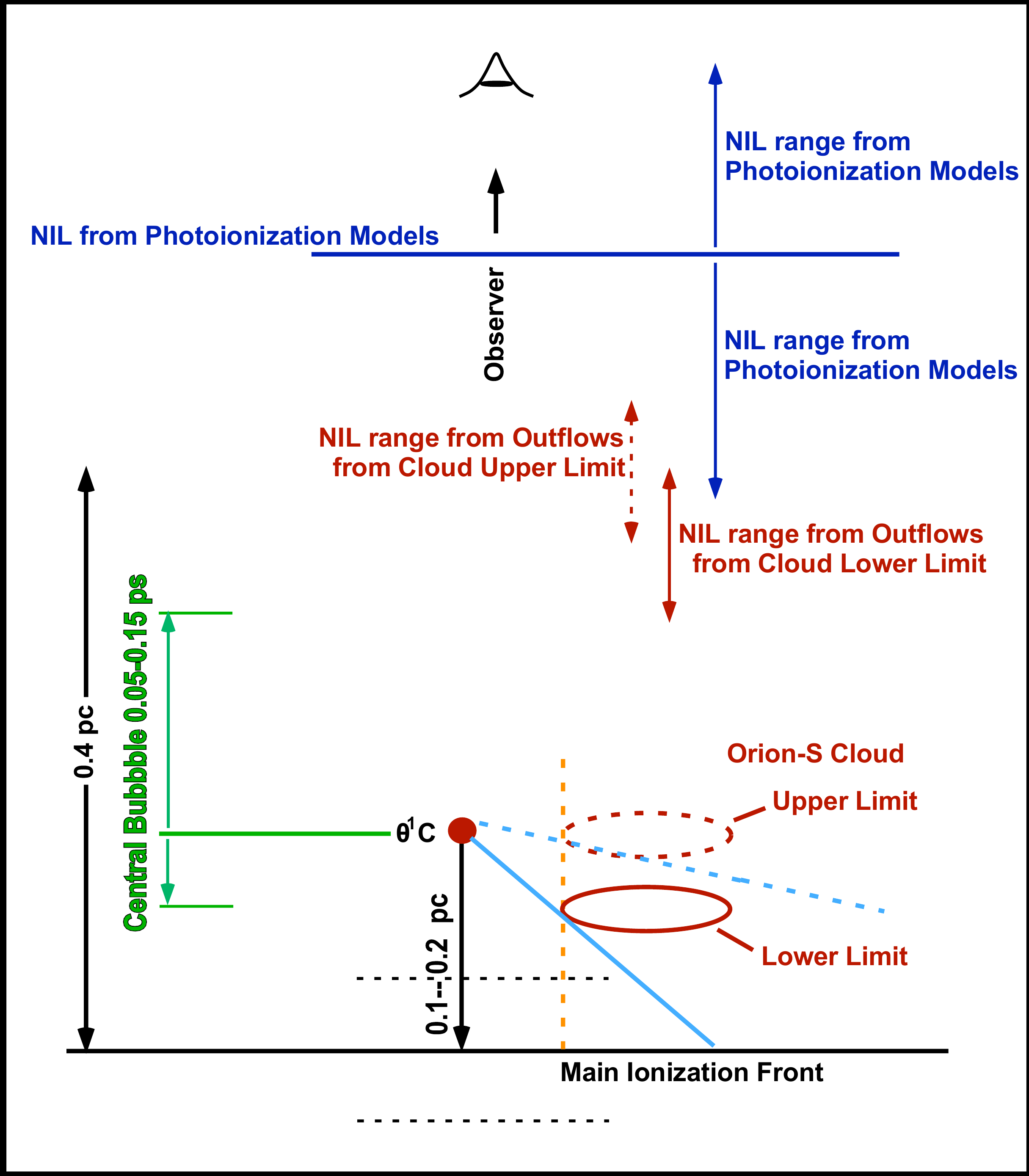}
 \caption{
 This working image shows the possibilities for the relative location of the MIF, \tC, the Orion-S Cloud, the boundary of the Central Bubble, and the NIL along a LOS through \tC\ (the red filled circle) as discussed in Section~\ref{sec:FinalDistances}. 
  }
\label{fig:fig7}
\end{figure}

Figure~\ref{fig:fig7} shows the results for derivation of differences in position along the line of sight. The distances are relative to \tC\ and different methods of deriving distances are color coded.

The \tC --MIF distance (black lines) in Figure~\ref{fig:fig7} was taken from \citet{ode17} and is based on the relative surface brightness of \nii\ and \oiii\ near \tC. \citet{ode17} adopted a distance of 0.15, which is smaller than obtained from surface brightness in hydrogen recombination lines. 
The dashed black lines show the possible range of values of this distance.

The radius of the High Ionization Arc  (green lines) is taken to represent the boundary of the Central Bubble, the range in distances reflects the fact that the Central Bubble may be non-circular and we have noted that the effects of the Central Bubble extend beyond the High Ionization Arc. 

The blue lines represent the \tC --NIL distance derived in Section~\ref{sec:refinement} from a photoionization model. The heavy blue line represents the center of the fitted region in Figure~\ref{fig:fig6}, while the dashed blue lines indicate the extreme values of the fitted region. 

Red is used to indicate features related to the Orion-S Cloud. 
The position of the Cloud is constrained laterally by the separation between \tC\ and the NW tilted boundary of the Cloud (33\arcsec ,0.061 pc , the orange dashed line in Figure~\ref{fig:fig7}). However, a range of values along the LOS for the Cloud can be derived from considerations 
of ionization of MIF material behind or shadowed by the Cloud, in addition to the Cloud's degree of ionization.

The Orion-S Cloud is sufficiently optically thick that it contains H$_{2}$CO and \hi , conditions that determine that it must be optically thick in Lyman continuum ionizing photons (LyC). This means that there will be a LyC shadow beyond the Orion-S Cloud. Where this shadow strikes the surface of the OMC, the MIF surface brightness will be low, but not zero since these regions will be illuminated by the diffuse LyC radiation field created by recombining gas. This shadowed region will be further from \tC\ as the distance between the Orion-S Cloud and the MIF is increased.

It is possible that there are regions of the MIF that are directly illuminated by \tC\ but hidden along the observer's LOS to the Orion-S Cloud. 
One sees H$_{2}$CO and \hi\ absorption lines against an ionized gas continuum in the direction of the Cloud. This means that a region beyond (further from the observer) but in alignment with the Cloud, is directly illuminated by \tC. This restricts the minimum MIF--Cloud distance to 0.1 pc, as shown in the solid red line ellipse and solid light-blue line in Figure~\ref{fig:fig7}.  

The Cloud could be even closer to the observer, allowing ionization of MIF material well beyond (away from \tC) the Cloud. 
However, an upper limit to this distance is when the \tC --MIF and Cloud-MIF distances are equal. Beyond that, the nearer (to the observer) side would not be ionized by \tC , in conflict with the fact that the near side of that Cloud is ionized. This upper limit is shown by the dashed red ellipse and dashed light-blue line in Figure~\ref{fig:fig7}. 
This means that the lower limit of the separation is about 0.1 pc and the upper limit about 0.15 pc.

In Section~\ref{sec:discNILandOriS} we established from outflows from the Orion-S Cloud that strike the NIL that their separation is 0.2 to 0.3 pc. Figure~\ref{fig:fig7} shows where this would place the NIL with respect to the two limits of the Cloud-MIF. 

The overall arrangement and scale of distances is satisfactory as it places the NIL at 0.4 pc in the same region from multiple approaches. It is attractive (Section~\ref{sec:BubbleNILlink}) as it satisfies the condition for the Central Bubble to be inside the NIL.

Of course all of this structure falls well within the distances \citep{abel19} of 2.0 pc for Veil component B and 4.2 pc for Veil component A.

\section{Discussion}
\label{sec:discussion}

\subsection{The Central Bubble}
\label{sec:bubble}

We interpret the High Ionization Arc to be the boundary in the plane of the sky of the Central Bubble surrounding the Trapezium stars. This cavity was originally proposed in a low spatial resolution (0\farcm9) radio map in H76$\alpha$ by \citet{pan79} but did not receive much attention until the optical study in \citet{ode09}, where its
properties were discussed in detail as the product of a hot, shocked, wind-blown cavity driven by the strong stellar wind from \tC\ \citep{how89,gagne05}. \citet{art12} modeled the properties of such a cavity while trying to explain the observed region of cool X-ray emitting gas found in the EON \citep{gud08}. She established that a central cavity of hot shocked gas was a natural product of the stellar wind and that an opening in it could produce the EON X-ray emitting material. X-ray emission should also occur in the region surrounding \tC , but that emission is absorbed by the foreground layers of the Veil.

Figure~\ref{fig:fig1} shows that our Large Samples 
in the NE-Region fall within the \oiii\ feature that defines the inner boundary of the High Ionization Arc.
 Immediately east of \tC\ the arc has a north-south diameter of 110\arcsec, corresponding to 0.2 pc. 
In contrast, the SW-Region data mostly come from a region where the High Ionization Arc is open, with the Orion-S Cloud in the direction 
of the opening. The SE-Region lines close but outside the SE boundary of the High Ionization Arc.

Following \citet{abel19} we attribute the \Vnew\ components (Section~\ref{sec:newblue}) to emission from the far (away from the observer) side and the \Vblue\ components (Section~\ref{sec:newblue}) to emission from the nearer side of the Central Bubble. 

In Figure~\ref{fig:fig5} we see that the far-side \Vnew\ components are most frequently found in the SW-Region, appearing there with equal frequency
in \Vnewnii\ and \Vnewoiii . This difficult to detect component is seen in \oiii\ in both the SE and NE-Regions, and only four times in \nii\ in the NE-Region.  The presence of \Vnewoiii\ outside of the High Ionization Arc can be used as an argument against this component occurring within the 
Central Bubble as presented in Section 5.2 of \citet{abel19}. However, the optical feature is likely to simply be the sharp edge of a more extended shell.

We also see in Figure~\ref{fig:fig5} that the \Vbluenii\ components are common in the NE and SW-Regions, while these Regions contain fewer \Vblueoiii\ features. The SE-Region is almost free of \Vblue\ components except for three Large Samples including \Vbluenii. The frequency and distribution of the \Vnew\ components argue for their being formed on the facing side of the Central Bubble.

In Section~\ref{sec:newblue} we found that \Vbluenii\ appears in 21 NE-Region Large Samples with an average of -6$\pm$3 \kms\ and in 11 SW-Region Large Samples with an average of -4$\pm$3 \kms, for a weighted average of -5$\pm$3 \kms. 
\Vblueoiii\ appears in five times in the NE-Region with an average of 1$\pm$4 \kms\ and ten times in the SW-region with an average of -1$\pm$5 \kms , for a weighted average of -1$\pm$4 \kms.

In Section~\ref{sec:newblue} we also found that \Vnewnii\ appears in four NE-Region Large Samples with an average of 37$\pm$12 \kms. It is much more abundant in The SW-Region Large Samples (appearing 26 times with an average of 33$\pm$2 \kms). The weighted average of both regions is \Vnewnii\ = 34$\pm$1 \kms.
\Vnewoiii\ appears 13 times in the NE-Region Large Samples (with an average of 27$\pm$5 \kms) and 20 times in the SW-Region (with an average of 27$\pm$8 \kms). It is also seen seven times in the SE-Region with an average of 27$\pm$2 \kms. The weighted average is \Vnewoiii\ = 27.0$\pm$5 \kms.

These velocities indicate that the boundaries of the Central Bubble are rapidly expanding. The side approaching the MIF is slowed by gas photo-evaporating from the MIF with the \Vnewnii~component moving away from \tC\ (assumed to be at 25 \kms\ \citep{sa05})  at 9 \kms\ and the \Vnewoiii\ component essentially at 2 \kms. 

The side approaching the observer is moving away from \tC\ at relative velocities 
of 30 \kms\ for \nii\ and 26 \kms\ for \oiii.

\subsection{Relation of the blue-shifted Central Bubble velocities and the NIL velocities}
\label{sec:BubbleNILlink}

The source of the blue-shift of the NIL is most likely to be the blue-shifted side of the expanding Central Bubble, which we saw in Section~\ref{sec:bubble} is approaching the observer at \Vbluenii\  = -5$\pm$3 \kms\ and  \Vblueoiii\ = -1$\pm$4 \kms. Given that the characteristic NIL velocities are \Vlownii\ = 5$\pm$3 \kms\ and \Vlowoiii\ = 8$\pm$ \kms, this means that the nearer side of the Central
Bubble are approaching the NIL at 10$\pm$4 \kms\ and 9$\pm$4 in \nii\ and \oiii\ respectively.

\subsection{Colliding features along the LOS}
\label{sec:collisions}

\begin{figure}
 \includegraphics
 [width=\columnwidth]
{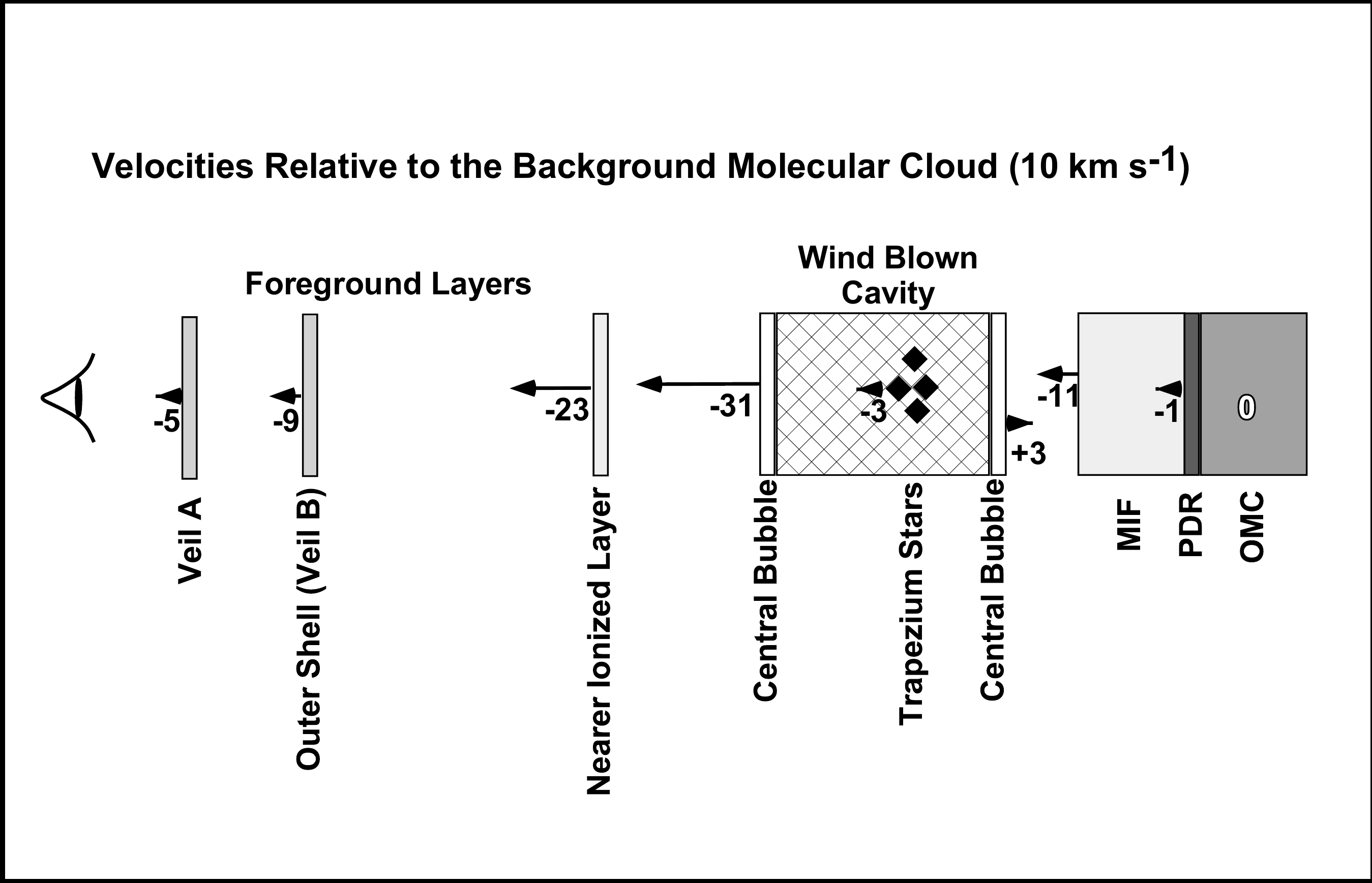}
 \caption{The positions (not to scale) and velocities relative relative to the OMC are shown and discussed in Section~\ref{sec:collisions}. This is an updated version of Figure~9 in \citet{abel19}. }
\label{fig:fig8}
\end{figure}

In our study of the many different features in the central Huygens Region we've found that everything is moving relative to one another.
We summarize these results in Figure~\ref{fig:fig8}. 

The closure time for the near side of the Central Bubble to the NIL is very short, in fact, we may already be seeing a case where the fast moving gas 
is pushing against the denser NIL. The closure time between the the NIL (at 0.4 pc and 6$\pm$2 \kms) and the Veil B component (at 2.0 pc and 19$\pm$1 \kms\ from \citet{abel19}  
is 1.2$\times$10$^{5}$ years, longer than the estimate of 30,000---60,000 years in \citet{ode18}.

\subsection{The nature of the scattering particles}
\label{sec:MIFnotNIL}

In Table~\ref{tab:DeltaVscat} we see that there is good agreement of the expected (isotropic scatterers) and observed velocity differences under the 
assumption of \Vmif\ being the source of the light that is backscattered. 

This difference in the \Vscat -\Vmif\ values can inform the question of the nature of the scattering properties of the particles in the PDR. If the backscattered light
was strongly concentrated back along the incoming beam, looking along a perpendicular tilted region would give a very weak \Vscat\ component and its velocity would \Vpdr, that is,
the same as the observed \Vmif, so that \Vscat -\Vmif\ = 0.  If the particles were isotropic scatterers, the observed \Vscat -\Vmif , would be the usual flat region value   2$\times$\Vevap. The fact that the data shown in Table~\ref{tab:DeltaVscat} are in good agreement with expectation is a strong argument that the scattering particles are nearly isotropic scatterers. However, the  closeness to true isotropy depends on the accuracy of our approximation that the velocity shift should be 2$\times$\Vevap.

\subsection{The putative relation of the \Vlow\ and \Vmif\ Components}
\label{app:Vcorr}

In \citet{ode18} it was argued that there is an approximately linear correlation between the \Vlow\ and \Vmif\ components.
The \Vlow\ component is always close to the \Vmif\ component (about 13 -- 18 \kms) and thus difficult to measure, because it is seen as a small signal on the blue shoulder of the strong \Vmif\ component. 
The conclusion in \citet{ode18} was that \Vmif -\Vlow\ is 18 \kms\ for \nii\ and 13 \kms\ for \oiii.

\citet{abel19} demonstrated that the linear correlation is very dependent on the mix of data employed. The \citet{ode18} study used data from both the Large Samples near the NE-Region and the results from lower S/N individual slits. They argue that the correlations becomes questionable if one only uses the Large Sample results. 
These cautions advise re-assessing the relation using the full dataset in the current study.

\begin{figure}
 \includegraphics
 [width=\columnwidth]
{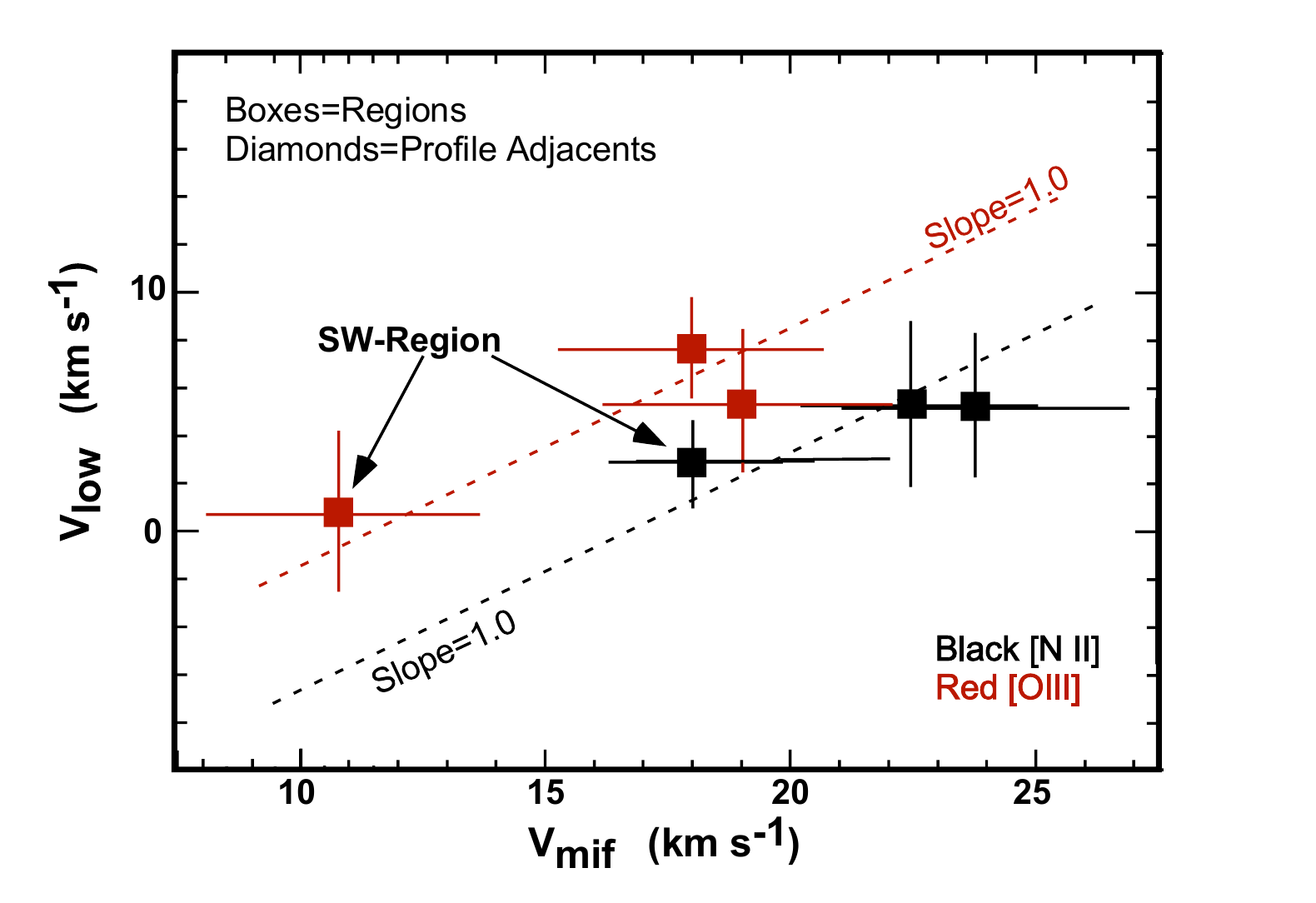}
 \caption{The possible correlation of \Vmif\ and \Vlow\ is illustrated in this figure. It is similar to Figures 4 and 5 of \citet{ode18} but more definitive as we now only use larger samples of higher S/N ratio in addition to showing probable errors. We identify the values for the SW-Region, which is
 affected by the Orion-S Cloud.
 }
\label{fig:fig9}
\end{figure}

We have used the Regions spectra to assess the likelihood of a correlation of \Vlow\ and \Vmif. These represent the highest S/N set of data available and they have identifiable probable errors. Figure~\ref{fig:fig9} was created  
using data from Table~\ref{tab:Regions}. In this figure we see several features.

The \nii\ values cluster around  \Vmifnii -\Vlownii\ = 16.5$\pm$3.0 \kms with an indication of a linear relation with a displacement of 16.5$\pm$2 \kms. 
A more apparent linear correlation of \Vlowoiii\ -- \Vmifoiii\ appears 
with a shift \Vmifoiii -\Vlowoiii\ = 11.5$\pm$2 \kms . The apparent correlations disappear if one disregards the SW-Region points. This would be justified since that region has anomalous values of many features and it is discussed in detail in Paper-II.

Rather than a correlation between the \Vmif\ and \Vlow\ velocities, an alternate view is that \Vmif\ varies because of differences in the inclination of the MIF, while \Vblue\ varies according to  
selection effects in analyzing the spectra. Previously we have argued that the apparent correlation of velocities are real \citep{ode18,abel19} but the current assessment indicates that they are associated with the intrinsic properties of the nebular lines and our method of analysis. 

We expect \Vmif\ to vary according to the inclination of the MIF that produces it. For a flat-on MIF \Vmif\ would be \Vpdr\ - \Vevap ,
20 \kms\ and 15 \kms\ for \nii\ and \oiii\ respectively. \Vmif\ for an edge-on MIF would be the \Vpdr\ value of 27 \kms.  
The \Vmif\ components shown in Figure~\ref{fig:fig9} are consistent with this expectation except for the SW-Region \Vmifoiii\ of 11$\pm$3 \kms. 

In Section~\ref{sec:Vlow} we saw that there is a host of data that argue for a constant \Vlow\ of 5$\pm$2 \kms, which is consistent with the \Vlow\ results shown in Figure~\ref{fig:fig9}, after consideration of the probable errors of their determination. The outlying sample is \Vlowoiii\ = 1$\pm$3 for the SE-Region. In a separate publication we will show that the \oiii\ emission in the SW-Region is highly irregular.

After consideration of the above material and the fact that the apparent \Vmif\ - \Vlow\ values lie close to the limits imposed by the FWHM of the \Vmif\ components, we must conclude that there is not a causal relation between the \Vmif\ and \Vlow\ velocities.

\subsection{Magnetic Fields in the Huygens Region}
\label{app:MagHR}

\begin{figure}
 \includegraphics
[width=\columnwidth]
 {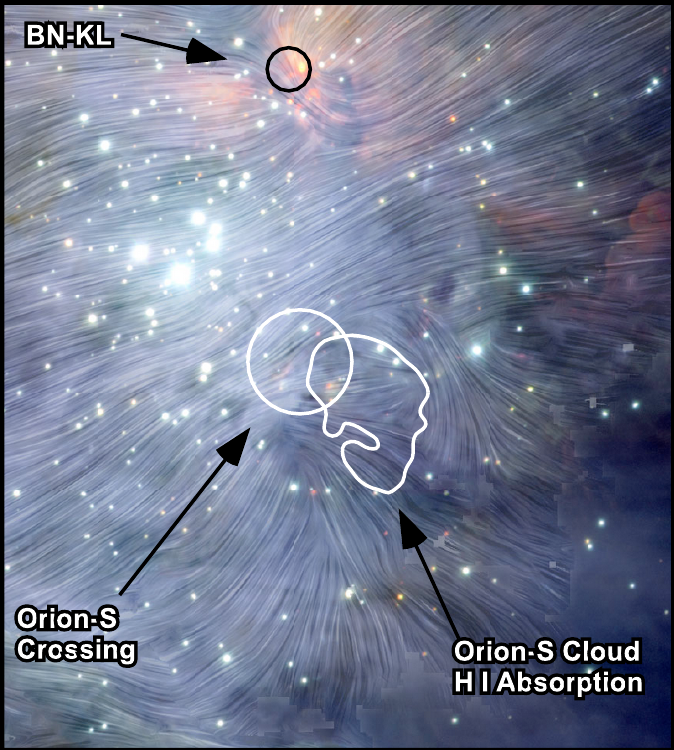}
 \caption{This 194\arcsec $\times$216\arcsec\ (0.36$\times$0.39 pc) infrared image of the Huygens Region (a portion of an European Southern Observatory press release 19 January 2001) has superimposed the magnetic field directions in the plane of the sky determined by \citet{chuss} from polarized 53 $\mu$m PDR dust emission at 5\farcs1 resolution(Astronomy Picture of the Day 27 February 2019). The location of the BN-KL imbedded complex of young stars is shown, as are the boundaries of the Orion-S Cloud as determined from \hi\ 21 cm absorption lines \citep{vdw13}, and the Orion-S Crossing.}
\label{fig:fig10}
\end{figure}

In a recent study from the Stratospheric Observatory for Infrared Astronomy (SOFIA) \citet{chuss} measured the polarization at multiple wavelengths of infrared continuum emission arising from dust in the Huygens Region. 
This radiation can arise from a heated imbedded region such as the active BN-KL region or from the high density PDR dust that is heated by radiation from the bright stars near the center of the Orion Nebula Cluster.  Their observations at 53 $\mu$m  provided the highest angular resolution (5\farcs1)
and at this wavelength they find B = 1000 \mG\ for the BN-KL Region and 261 \mG\ for a quiescent region. The latter number is much larger than the characteristic average of 56 $\mu$G) found in the foreground Veil A component by \citet{tom16}. 

\subsubsection{Magnetic Field directions near BN-KL and the Orion-S Cloud}
\label{app:MagHR1}

Figure~\ref{fig:fig10}
is a cropped segment of an infrared image with the Chuss et al. results super-imposed.
It shows that in this area the direction of the magnetic field at 53 $\mu$m usually varies only slowly with position with the exception of the BN-KL region and the Orion-S Cloud.The variations in the BN-KL region are discussed in detail in \citet{chuss}; but, not the Orion-S Cloud properties which are emphasized in our discussion.

The region around the Orion-S Crossing and the Orion-S Cloud has properties obviously relevant to the local structure. 
Figure~1 of Chuss shows that the magnetic field is weaker than average in the inner parts of what we designate as the SW-Region and our Figure~\ref{fig:fig10} shows that the large-scale direction of the magnetic field in the SW-Region is perpendicular to that in the center of the Huygens Region. The local peak surface brightness at 53 $\mu$m occurs near the center of the Crossing. Evidently the PDR producing the 53 $\mu$m radiation is of a very different geometry than the surrounding nebula. In addition, the direction and lower strength of the field then extends into the SW-Region. This strengthens our argument that this region is different from other parts of the Huygens Region and that the cause of the difference lies with the Orion-S Cloud and its NE corner that is strongly illuminated by \tC. These are elaborated upon in Paper-II.

\subsubsection{A connection of different regions through Ambipolar Diffusion ?}
\label{app:MagHR2}

Given the wealth of information on the Orion complex, it is possible to compare the magnetic field and density for different regions of Orion. 
\citet{cru} found, from 27 Zeeman observations of molecular clouds, a relationship between magnetic field strength and density, B$\propto$n$^{0.47}$.  This result is consistent with ambipolar diffusion-driven star formation, where the magnetic field-density relation would be given by B$\propto$n$^{0.5}$.
Given the known magnetic field and density in the neutral layers of the Veil, we can calculate the expected density in BN-KL and the HII region based on the relation from \citet{cru}.  For Veil component A, B$\simeq$56 $\mu$G \citep{tom16} and the density is 10$^{2.4}$ cm$^{-3}$ \citep{abel16}.  Using these values, the magnetic field of BN-KL yields a density of 10$^{5.1}$cm$^{-3}$, and the \hii\ region gives a density of 10$^{3.8}$cm$^{-3}$.  These densities are very close to the densities derived from PDR modeling in \citet{mor16} of 10$^{5.3}$cm$^{-3}$ (for BN-KL) and of 10$^{3.8}$ cm$^{-3}$ (for the \hii\ region) \citep{bal91}. This argues for the possibility that ambipolar diffusion is a physical process connecting multiple regions of Orion, such as the Veil and BN-KL, even though their physical separation exceeds 2 parsecs.

\subsection{A recent study of the Veil B \Cii\ component}
\label{sec:pabst}

In a recent paper \citet{pabst}  reported mapping a 1$\arcdeg$  field around the Huygens Region in the \Cii\  158 $\mu$m line at 0.2 \kms\ and 16\arcsec\ resolution.
They identified a curved shell of about 2500\arcsec\ (41\farcm7 ) diameter (4.6 pc at 383 pc distance) and a maximum expansion velocity towards the observer 
of 13 \kms. We'll refer to this as the Outer Shell. They then modeled this feature as a 2 pc diameter bubble and explained its structure and dynamics as the result of the hot gas \citep{gud08} created by reverse shocks from the stellar wind sweeping up surrounding gas and forming a shell. 

The discovery of the Outer Shell is an important step in understanding the EON. A draft of a longer paper expanding on \citet{pabst} shows that the Outer Shell velocity near the Trapezium is 19 \kms\ while the component associated with the background PDR is at 28 \kms.

This portion of the Outer Shell in the direction of the Huygens Region has previously been studied, before the recognition of the Outer Shell. It was discovered in the 21-cm \hi\ absorption line
study of \citet{vdw89} as component Veil B at 19.4 \kms,  seen in \Htwo\ ultraviolet absorption lines at 19.5$\pm$0.7 \kms\ \citep{abel16} , seen in \Caii\ and \Nai\ optical absorption lines at 18.3 \kms\ and 19.8 \kms\ respectively \citep{ode93}, and at 19$\pm$2 \kms\ in \Cii\  \citep{goi15} using the same emission line (0.2 \kms\ and 11\farcs4 resolution). These velocities were summarized in Table~2 of \cite{abel19}. The weighted average of the components is 
19.2$\pm$0.5 \kms.

There is an open question of about how to reconcile the results of this study (which argues that the stellar wind escapes the Central Bubble only to the SW) and the fact that the Outer Shell extends across and slightly north of the Trapezium.



 \section{Conclusions}
    \label{sec:conclusions}
    
$\bullet$ \tC\ is surrounded by a wind-blow Central Bubble open to the SW and about 110\arcsec\ (0.21 pc) north to south width at the star.

$\bullet$ There is a layer of ionized gas (the NIL) extending across the Huygens region. At 0.4 pc from \tC, it lies just outside the Central Bubble.

$\bullet$ The dust particles in the PDR are isotropic back-scatterers.

$\bullet$ A previously proposed relation between \Vmif\ components arising from the MIF and \Vlow\ components arising from the NIL is shown to be unlikely. Its appearance is primarily due to the difficulty of measuring a weak line on the shoulder of the \Vmif\ component.

$\bullet$ The Veil B components, seen in \hi\ and multiple ions appear to be part of an Outer Shell discovered in high velocity resolution mapping of the EON.

$\bullet$ Our SW-Region sample has unusual \Vmifoiii\ properties and is located in a region where the PDR's magnetic field orientation and strength is different from other parts of the Huygens Region (with the exception of the area near the BN-KL objects.

 \section*{acknowledgements}
We are grateful to C. H. M. Pabst of Leiden University for sharing a draft of her paper expanding on \citet{pabst}, discussions on the Outer Shell, and her comments on a draft of this paper.
The observational data were obtained from observations with the NASA/ESA Hubble Space Telescope,
obtained at the Space Telescope Science Institute, which is operated by
the Association of Universities for Research in Astronomy, Inc., under
NASA Contract No. NAS 5-26555; the Kitt Peak National Observatory and the Cerro Tololo Interamerican Observatory operated by the Association of Universities for Research in Astronomy, Inc., under cooperative agreement with the National Science Foundation; and the San Pedro M\'artir Observatory operated by the Universidad Nacional Aut\'onoma de M\'exico. GJF acknowledges support by NSF (1816537, 1910687), NASA (ATP 17-ATP17-0141), and STScI (HST-AR- 15018). 

\appendix

\section{The visibility of the \Vlow\ component is determined by the FWHM of the \Vmif\ component}
\label{sec:FWHM}

It is prudent to examine if the \Vlow\  results are due to the manner in which the spectra were analyzed. In Appendix A of \citet{ode18} it was illustrated how artificially created spectra closely resemble the observed spectra, using the 
\nii\ line for the comparison. However, this illustration does not critically test the visibility of the \Vlow\ component.
It is intuitively obvious that if the \Vmif\ component is broad, it will be difficult to find a weak \Vlow\ component on its blue shoulder. The tests we describe below are intended to quantitatively evaluate the limits of detection of the \Vlow\ component.

A series of model spectra were created using varying relative strengths and displacements of the \Vlow\ components.
These were created by adding the \Vlow\ components to a fixed spectrum composed of a \Vmif\ component with a variable FWHM plus a 
\Vscat\ component of FWHM = 22 \kms\ and displaced 20 \kms\ to the red. These were called the RED spectra. A series of 
RED spectra were created with \Vmif\ FWHM values of 10, 12, 14, and 16 \kms\ . For each of these a series of \Vlow\ spectra were
added, with the same FWHM for \Vlow\ and varied assumed values of \Slow /\Smif. This process should test whether a grouping of \Vlow\ - \Vmif\ values as shown in Figure~\ref{fig:fig9} are 
a product of the analysis of the data, rather than revealing a true correlation.

\begin{figure}
 \includegraphics
[width=3.0in]
{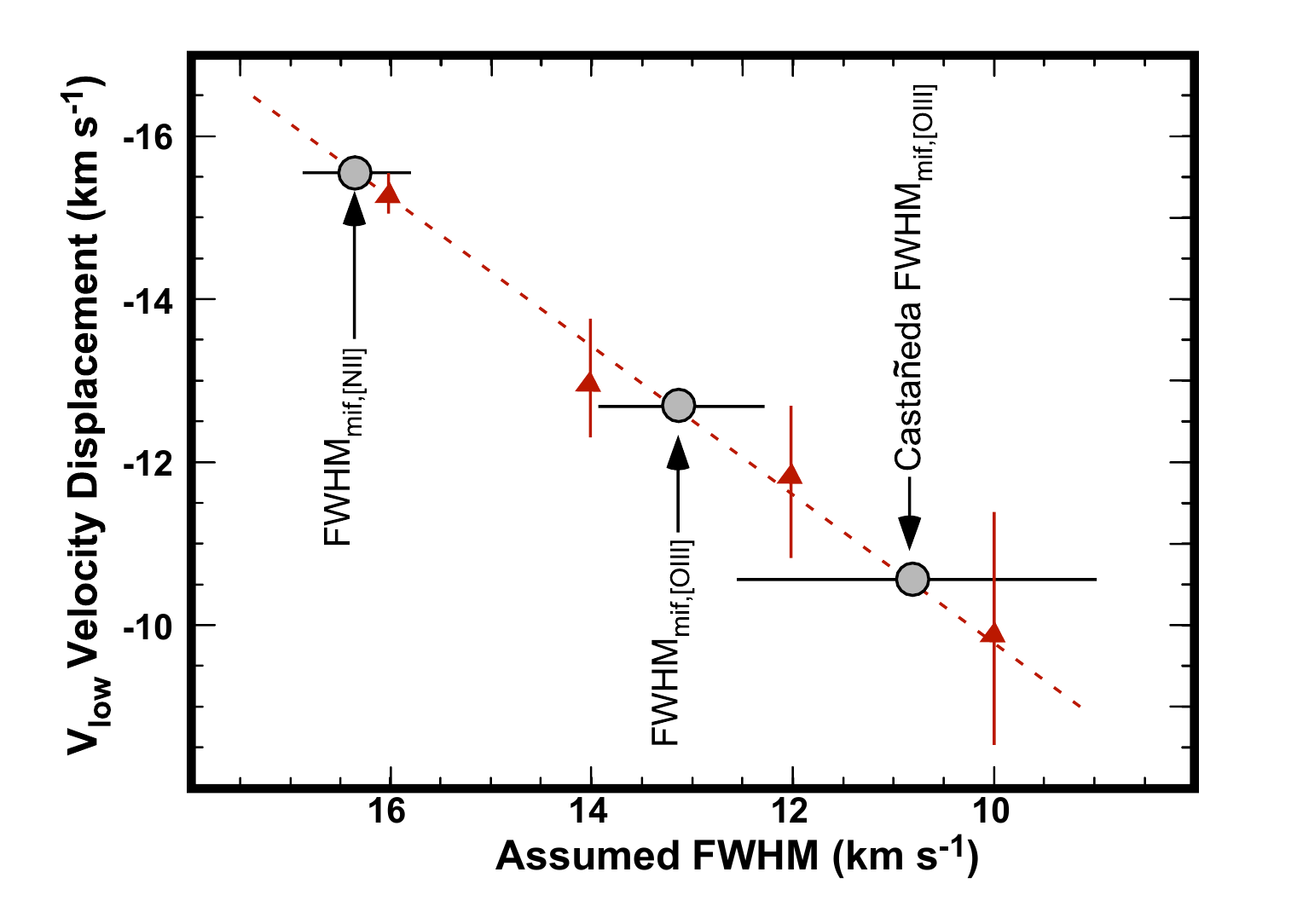}
 \caption{The red triangles show the minimum measurable velocity displacement for \Vlow\ components as a function of the FWHM of the \Vmif\  components (\Vlow\ is assumed to the the same) for the models explained in Section~\ref{sec:FWHM}. The dispersion in the measured \Vlow\ displacements is from five independent deconvolution using 'splot' for each limiting detection spectrum. \Vlow\ components above the dashed red line should be clearly measured in actual spectra of a specific FWHM. The large grey filled circles indicate the average FWHM of the MIF component
 in our study and in the higher resolution study of \citet{hoc88} of \oiii. }
\label{fig:fig11}
\end{figure}

It was found that at a fixed FWHM and diminishing values of \Slow /\Smif\ a point was reached where the \Vlow\ component had disappeared from visibility in the blue wing of the composite spectrum. That point was set at the same level of visibility employed in the measurement of the nebular spectra. Larger displacements than this limit would have been clearly identified and measured. Below that point, no \Vlow\ component would have been measured. The results are shown in Figure~\ref{fig:fig11} as red triangles.  

The limiting \Vlow\ values in Figure~\ref{fig:fig11} fall along an indistinguishably linear relation (shown as a red dashed line) 
with the limiting velocity being slightly less than the assumed FWHM. This figure also shows the average FWHM of the 
\Vmifnii\ and \Vmifoiii\ values in the Regions. When those FWHM values are shifted to lie on the 
dashed red line, they should indicate the minimum displacement of the \Vlow\ component that can be measured. 

Nearly all of the nebular spectra \Vmifnii\ components (FWHM=16.4$\pm$0.6 \kms) had detectable \Vlownii\ components, while only some of the \Vmifoiii\ components did (FWHM=13.2$\pm$0.8 \kms). 

At the average FWHM of the \Vmifnii\ components the expected minimum \Vlownii\ shift would be -15.5$\pm$0.5 \kms. At the average of the \Vmifoiii\ components with \Vlowoiii\ components the expectation is  -12.7$\pm$0.8 \kms.  As discussed in Section~\ref{sec:FWHM} the observed \Vlownii\ components lie in a region relatively unaffected by the process of its identification, but the \Vlowoiii\ components are. In the latter case, we draw on the results of the higher resolution study of \oiii\ by \citet{hoc88}. 

\section{Revisions to the LOS Table presented in Abel et al. (2019).}
\label{app:revisions}

Table~2 of \citet{abel19} summarized the velocities of various velocity components, using data from their study 
and previously published material. Our study has doubled the number of spectra and we have revised portions of that table, creating Table~\ref{tab:velocities}. 
We now use the term Nearer Ionized Layer instead of the previous Ionized Component I.
\begin{table*}
\caption{Velocity Systems*}
 \label{tab:velocities}
 \begin{tabular}{ccc}
 \hline
 \hline
\colhead{Designation} &
\colhead{Component and Velocity* (\kms) } &
\colhead{Source}\\
\hline
Nearer Ionized Layer & (6$\pm$2)             & Weighted Average\\
---                & \Vlownii\ (5$\pm$3)             & All three Regions\\
 ---               & \Vlowoiii\ (8$\pm$3)            & \citet{hoc88}\\
 ---              & \oii\ (system B, 3.1)                    & \citet{jones92}\\
 ---                & \siii\ (system B, 9.9)                     & \cite{wen93}\\
 ---                & \heI\ (absorption, 2.1$\pm$0.6)&\citet{ode93}\\
 ---                & \Piii\ (absorption, 4.9$\pm$3.0) & \citet{abel06}\\
 ---                & \siiia\ (absorption,4.5$\pm$0.9)  & \citet{abel06}\\
 ---                 & \Caii ** (absorption, 7.5)              &\citet{ode93}\\
 ---                 & \Nai *** (absorption, 6.0)               & \citet{ode93}\\
Nearer Central Bubble     & \Vbluenii  (-5$\pm$3) &Two Regions\\
---                   & \Vblueoiii\ (-1$\pm$5) & SW-Region\\
 Cluster Stars & Stellar Spectra (25$\pm$2)         & \cite{sa05}\\
 Further Central Bubble      & \Vnewnii\ (35$\pm$3)            & Two Regions\\
 ---                 & \Vnewoiii\ (27$\pm$4)           & All three Regions\\ 
 Main Ionization Front (MIF)& \oi\ (27$\pm$2)    & \citet{ode92a}\\
 ---                 &  \oii\ (18$\pm$1)           &\citet{adams44,jones92}\\
 ---               &  \nii\ (22$\pm$2)                  &All three Regions\\
 ---                 & \siii\ (20$\pm$4)                    &\cite{wen93}\\
 ---                 &  \oiii\ (17$\pm$3)                  &All three Regionss\\
 ---                 & H8+H12 (17$\pm$2)                &\citet{ode93}\\
 ---                 & H41$\alpha$ (16$\pm$2)        &\citet{goi15}\\
 ---                 & \Heplus\ (17$\pm$2)                & \citet{ode93}\\
  Backscattered MIF & \nii\ (40$\pm$2)                & All three Regions\\
 ---                  & \siii\ (36$\pm$4)                 & \citet{wen93}\\
 ---                  & \oiii\ (39$\pm$1)                 & All three Regions\\
 \hline
\end{tabular}\\

~**\Caii\ 393.4 nm.
***\Nai\  589.6 nm.
\end{table*}

\clearpage
\end{document}